\documentclass[aps,twocolumn,floatfix,showpacs]{revtex4-1}
\usepackage[utf8]{inputenc}

\usepackage{amsfonts}
\usepackage{amsmath}
\usepackage{amssymb}
\usepackage{graphicx}
\usepackage{bm}
\usepackage{color}

\def\be{\begin{equation}}     
\def\ee{\end{equation}}
\def\ba{\begin{eqnarray}}
\def\ea{\end{eqnarray}}
\def\<{\left<}
\def\>{\right>} 
\def\({\left(}
\def\){\right)} 
\def\[{\left[}
\def\]{\right]} 
\def\cd{\mathrm{cd}}
\def\blue{{}}

\begin{document}

\title{Electroviscous drag on squeezing motion in sphere-plane geometry}
\author{Marcela Rodr\'iguez Matus$^{1*}$, Zaicheng Zhang$^{1*}$, Zouhir Benrahla$^{1*}$, Arghya Majee$^{2}$, Abdelhamid Maali$^{1}$, Alois Würger$^{1}$}
\affiliation{$^1$Université de Bordeaux $\&$ CNRS, Laboratoire Ondes et Matière d’Aquitaine,  33405 Talence, France\\
$^2$Max Planck Institute for Intelligent Systems, Stuttgart, Germany $\&$ IV. Institute for Theoretical Physics, University of Stuttgart, Germany}

\begin{abstract}
	Theoretically and experimentally, we study electroviscous phenomena resulting from charge-flow coupling in a nanoscale capillary.  Our theoretical approach relies on Poisson-Boltzmann mean-field theory and on coupled linear relations for charge and hydrodynamic flows, including electro-osmosis and charge advection. With respect to the unperturbed Poiseuille flow, we define an electroviscous coupling parameter $\xi$, which turns out to be maximum where the film height $h_0$ is comparable to the \blue{Debye} screening length $\lambda$. We also present dynamic Atomic Force Microscope (AFM) data for the visco-elastic response of a confined water film in sphere-plane geometry; our theory provides a quantitative description for the electroviscous drag coefficient and the electrostatic repulsion as a function of the film height, with the surface charge density as the only free parameter. Charge regulation sets in at even smaller distances.
\end{abstract} 

\maketitle

\section{Introduction}

Solid surfaces in contact with water are mostly charged, resulting 
in intricate interactions of the diffuse layer of counterions with  liquid flow along the solid boundary \cite{Israelachvili1991,israelachvili1996role,Lyklema1995}. Charge-flow coupling is at the origin of various electro-kinetic and  electric-viscous effects \cite{stone2004engineering}. Besides  classical applications of capillary electrophoresis ranging from microfluidics to medical analysis, recently AC charge-induced electro-osmosis has been used for the assembly of active materials from micron-size colloidal building blocks \cite{yan2016reconfiguring}, 
surface osmotic effects have been discussed in view of energy  
applications and desalinization of sea water \cite{Marbach2019}. 

\blue{
The underlying physical mechanisms operate on the scale of the Debye screening 
length \cite{bocquet2010nanofluidics}, which is of the order of a few tens of nanometers.}
Following the derivation of the electroosmotic coefficient by 
Helmholtz \cite{Helmholtz1879} and Smoluchowski \cite{Smoluchowski1903}, 
electrokinetic effects have been extensively studied in the limit 
of thin double layers, where  the  screening  length is  much 
smaller than the depth of the liquid phase. Thus 
Bikerman \cite{Bikerman1933} and Dukhin \cite{Lyklema1995} 
derived the surface contribution to the electric conductivity 
of an salt solution, and H\"uckel \cite{Hueckel1924} and 
Henry \cite{Henry1931} showed the colloidal electrophoretic 
mobility to depend on the ratio of particle size and screening 
length. Gross and Osterle studied charged membranes separating 
two electrolyte solutions at different pressure and 
electro-chemical potentials, and numerically calculated the 
transport coefficients of nanopores comparable to the screening 
length \cite{Gross1968}.

\begin{figure}[b]
		\includegraphics[width=\columnwidth]{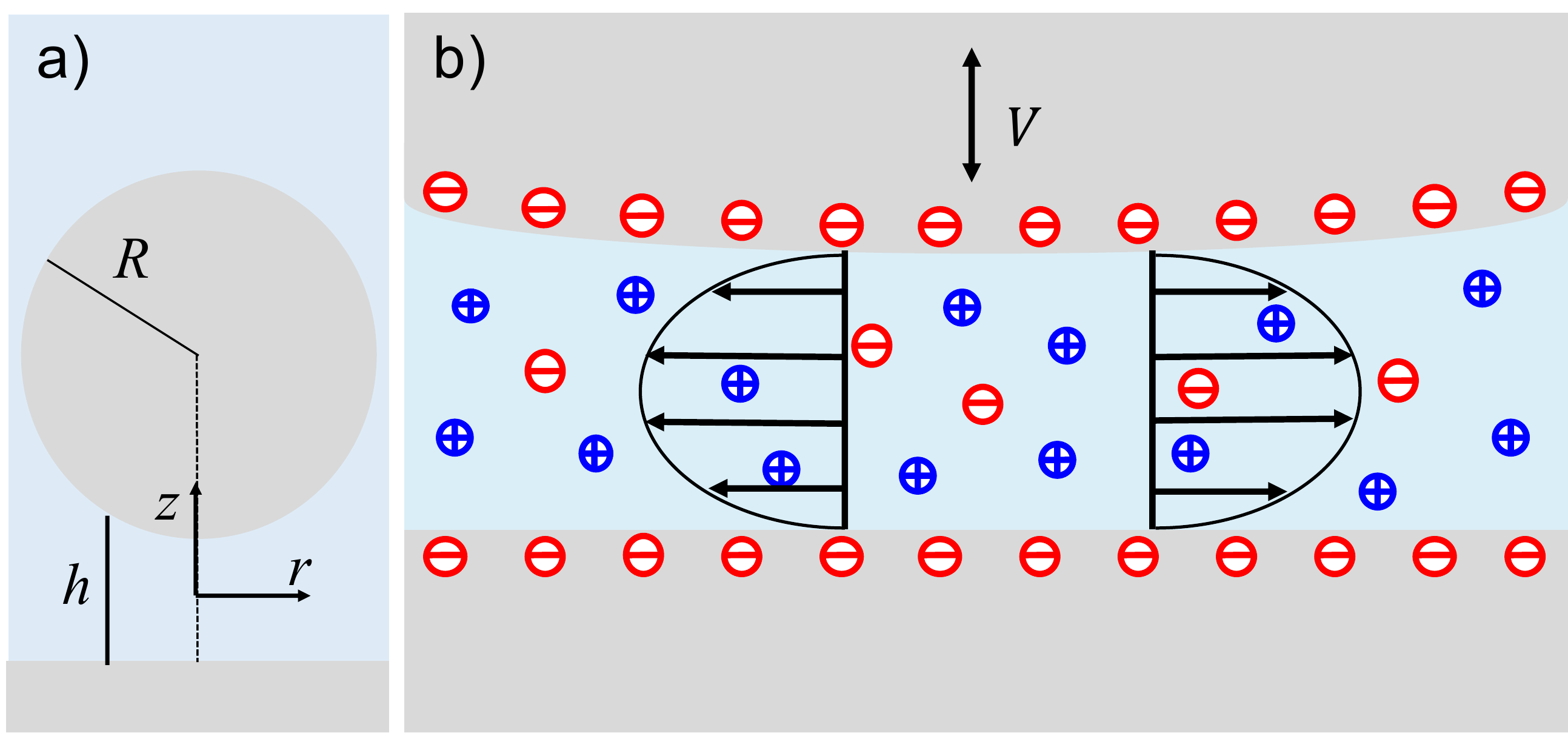}
			\caption{\blue{Schematic view of charge-flow coupling in sphere-plane geometry. a) A colloidal sphere of radius $R$ is placed above a solid surface. The film height $h$ varies with the radial coordinate $r$ and takes its minimum value $h_0$ at $r=0$; this distance satisfies $h_0\ll R$. b) The sphere vertically vibrates with velocity $V(t)$. This squeezing motion induces a radial Poiseuille flow in the confined water film containing mobile ions of either sign. }}
	    \label{fig:1}
\end{figure}

\blue{ 
Prieve and collaborators studied charge effects on the motion of a colloidal sphere moving close to a solid surface \cite{Alexander1987,Bike1990,Bike1992,Bike1995a,Bike1995b}. For a particle sliding parallel to the surface at velocity $V$, they observed a normal lift force proportional to $V^2$. This dependence suggests as underlying mechanism the Maxwell stress $\varepsilon E^2$, with the permittivity $\varepsilon$ and the parallel electric field arising from the streaming potential, $E\propto V$ \cite{Bike1990}. Yet the measured lift force \cite{Bike1995a} by far exceeds the calculated value \cite{Bike1995b}; this discrepancy has not been elucidated so far.} 

\blue{
Quite a different situation occurs for the squeezing motion of a colloidal sphere vibrating in normal direction with a sinusoidal displacement $Z(t)$, as shown schematically in Fig. \ref{fig:1}. The velocity $V=dZ/dt$ is by orders of magnitude smaller than that of sliding motion, resulting in a negligibly weak electrokinetic lift. For uncharged surfaces, the only force at work is the hydrodynamic drag $-\gamma_0V$ with coefficient $\gamma_0$. The presence of electric 
double layers gives rise to several electrokinetic forces,
\begin{equation}
    K-kZ-\gamma V,
    \label{eq:0}
\end{equation}
where the electrostatic repulsion $K$ is well known from static Atomic Force Microscope (AFM) experiments \cite{raiteri1996measuring}. For a mechanically driven system as in Fig. \ref{fig:1}, the dynamic response consists of a restoring force $-kZ$ with an effective spring constant $k$ and an enhanced drag coefficient $\gamma$, due to the coupling of the charged diffuse layers to the radial flow profile  \cite{Liu2015,Liu2018}. 
Bike  and Prieve calculated the charge contribution $\gamma-\gamma_0$ for the case where the sphere-plane distance $h_0$ is much larger than the Debye screening length  $\lambda$ \cite{Bike1990}. 
Subsequent numerical studies discussed the enhancement 
factor  for both narrow and wide channels, and found 
a maximum to occur  at $\lambda/h_0\approx1$ 
\cite{Chun2004,Zhao2020}. The first unambiguous 
experimental observation of the electroviscous 
effect was reported very recently by Liu et al, 
who performed dynamic AFM experiments in weak 
electrolyte solutions \cite{Liu2018}. }

\blue{The present work intends to clarify whether charge-flow coupling accounts quantitatively for the electroviscous drag on the squeezing motion illustrated in Fig. \ref{fig:1}. 
Section II provides a brief reminder of Poisson-Boltzmann  theory, and the static repulsive force $K$ and the spring constant $k$. In Section III we develop the formal apparatus for charge-flow coupling, relying on Onsager's phenomenological relations for generalized fluxes and forces, without resorting to the linearization approximation in the  electroviscous coupling parameter. We derive the 
electroviscous drag coefficient $\gamma$ in terms of the Onsager transport coefficients $L_{ij}$. In Section IV we compare analytical approximations for the limiting cases of narrow and wide channels with the numerical computation. Section V is devoted to a discussion of the 
effect of charge regulation on both electrostatic and electroviscous properties. In Sect. VI we present 
dynamic-AFM measurements and compare with our theoretical 
findings. }

\section{Electrostatics}

{}{
Here we briefly discussed the electrostatic properties in the absence of external driving. Solid materials in contact with water in general carry surface charges. Due to electrostatic screening, the released counterions are confined in a diffuse layer of charge density $\rho$, which is related to the electrostatic potential $\psi$ through Gauss's law  
  \be
    \nabla^2\psi = -  \frac{\rho}{\varepsilon}.
  \ee
In the framework of Poisson-Boltzmann mean-field theory the concentrations 
of monovalent ions  read $n_\pm = n_0 e^{\mp e\psi/k_BT}$, where 
the bulk value $n_0$ corresponds to dissolved salt, to carbonic acid absorbed from air or to the dissociation of water. The resulting expression for the charge density 
\be
\rho = e(n_+ - n_-) = 2 e n_0 \sinh\frac{e\psi}{k_BT} 
\ee
then closes Gauss's law. 
}

\subsection{1D Poisson-Boltzmann theory}

\blue{
This work deals with thin films as in Fig. \ref{fig:1}, where the minimum  height is much smaller than the radius of the vibrating sphere, $h_0\ll R$. Then electrostatic and hydrodynamic properties are relevant in the lubrication area only, which corresponds to the range where the radial coordinate $r$ 
takes values much smaller than $R$ and where the height $h(r)$ of the aqueous film is a slowly varying function of $r$. For notational convenience we define the  origin of the
vertical coordinate $z$ such that the solid boundaries are at $z=\pm h/2$. }
 
 \blue{
 Throughout this paper we assume a homogeneous surface charge and use the 1D Poisson-Boltzmann equation 
where $\psi$ and $\rho$ depend on the vertical coordinate $z$ only,
\begin{align}
  \frac{e}{k_BT}\frac{d^2\psi}{dz^2} 
    = \lambda^{-2} \sinh\frac{e\psi}{k_BT},
      \label{eq:1}
\end{align}
Here we introduce two characteristic length scales, the Debye screening length 
 \begin{align}
    \lambda = \frac{1}{\sqrt{8\pi n_0 \ell_B}}
      \label{eq:1b}
 \end{align}
which gives the thickness of the diffuse layer in an electrolyte solution \cite{Andelman2006}, and the Bjerrum length
\begin{align}
    \ell_B=\frac{e^2}{4\pi\varepsilon k_BT}
\end{align}
which gives the distance where the electrostatic interaction of two elementary charges is equal to the thermal energy. Typical values in water are $\lambda=1...300$ nm and $\ell_B=0.7$ nm}

For fixed surface charge density $e\sigma$, the potential satisfies
the boundary condition 
\begin{align}
  \frac{e\sigma}{\varepsilon} 
  = \mp \left.\frac{d\psi}{dz}\right|_{z=\pm h/2}. 
\label{eq:2}
\end{align}
 For fixed surface potential one has $\psi(\pm h/2) = \zeta$. Note 
 that the potential $\psi(z)$ and its surface value $\zeta$ depend 
 on the film height $h$ and thus on $r$.

\subsection{Disjoining pressure and repulsive force}

For the sake of notational simplicity we assume a symmetric 
system with  the same charge density $\sigma$ on the two 
opposite surfaces. Then the disjoining pressure is given by 
the excess osmotic pressure of the mobile ions at $z=0$, which reads 
$\Pi=(n_m - 2n_0)k_BT$. With the excess number density 
$n_m=2n_0\cosh(\psi(0)/k_BT)$, one readily finds
\begin{align}
  \Pi = 2 n_0 k_BT \left(\cosh\frac{e\psi(0)}{k_BT}-1\right). 
  \label{eq:Pi}
\end{align}
The dependence of the osmotic pressure on the film height $h$ 
arises from the potential $\psi(z=0)$ \cite{Andelman2006}. 
At distances $h$ larger than the screening length  $\lambda$, this 
potential vanishes, and so does the disjoining pressure. 

The repulsive force $K$ between the two surfaces, is obtained as the surface integral the 
osmotic pressure. The film height being much smaller than 
the curvature radius, we use the Derjaguin 
approximation \cite{Derjaguin1934}. For distances much 
smaller than the radius of the oscillating sphere, the 
 height of the water film $h=h_0+R-\sqrt{R^2 - r^2}$
is well approximated by
\begin{align}
    h(r) = h_0 + \frac{r^2}{2R},\;\;\;\;\; (r\ll R).
    \label{eq:19}
\end{align}
Writing the surface element as $dS=2\pi drr = 2\pi Rdh$, 
one readily obtains
\begin{align}
 K(h_0) = \int dS \Pi = 2\pi R\int_{h_0}^\infty dh\Pi(h).
 \label{eq:60}
\end{align}
The disjoining pressure gives rise to a static restoring force $-kZ$, with spring constant
\begin{align}
 k(h_0) = -\frac{dK}{dh_0} = 2\pi R \Pi(h_0).
 \label{eq:61}
\end{align}
The discussion and numerical evaluation of the force $K$ and the 
rigidity $k$ are postponed to  Sect. \ref{section:CR} below.

\section{Charge-flow coupling: Formal apparatus} 

\blue{
Here we derive the formal expression for the electroviscous drag coefficient $\gamma$ defined in \eqref{eq:0}.  Resorting to lubrication approximation, we give the coupled hydrodynamic and charge flows in radial direction, which are imposed by the mechanical driving, as illustrated in Fig. 1. Then we derive expressions for the hydrodynamic pressure and the resulting drag force.}

We consider charged surfaces in sphere-plane geometry, 
in contact with an electrolyte solution, as shown 
schematically in Fig. \ref{fig:1}. The vertical distance 
varies with time according to $h_0+Z(t)$, with a small 
sinusoidal amplitude $|Z|\ll h_0$ and frequency $\omega$, 
resulting in the velocity $V = dZ/dt$. Experimentally, 
this is realized by a vibrating sphere of radius $R$ 
mounted on the cantilever of an AFM. 

\subsection{Lubrication approximation}

The vertical oscillation modulates the hydrodynamic pressure 
$P$ in the film and imposes a flow $J_V$. For an incompressible 
fluid, there is a simple geometrical relation between the 
vertical velocity $V$ of the cantilever and the volume flow 
carried by the radial fluid velocity $v$, 
\blue{
\begin{align}
 \pi r^2 V = 2\pi r J_V = 2\pi r \int_{-h/2}^{h/2} dz v(z,r).
 \label{eq:12}
\end{align} 
Note that the height  $h(r)$ varies with the radial position $r$ according to \eqref{eq:19}.}

The fluid mechanical problem simplifies significantly when 
resorting  to the lubrication approximation \cite{Happel1963}. 
In the range of validity of Eq.~\eqref{eq:19}, the vertical 
component of the velocity field is negligible, and the 
radial component $v$ obeys a simplified 
Stokes equation, 
\begin{align}
\eta\partial_z^2v = \partial_r P - \rho E,
\label{eq:9}
\end{align}
with the  viscosity $\eta$ and where only the vertical 
component of the Laplace operator $\nabla^2 v$ has been 
retained. The right-hand side comprises the radial pressure gradient $\partial_r P$ and the force exerted by a radial electric field $E$ and the charge density $\rho$ of the diffuse layer.

\subsection{Non-equilibrium fluxes and forces}

Using the Derjaguin approximation, the electrostatic properties can be calculated from the 1D Poisson-Boltzmann equation \eqref{eq:1} with slowly varying gap height $h(r)$. Yet this equilibrium state is perturbed by charge-flow coupling. Indeed, advection of counterions by the radial velocity $v$, results in a radial charge distribution and an electric field $E$. Through the electro-osmotic force $\rho E$ in \eqref{eq:9}, the field backreacts on the flow properties.  

For an axisymmetric geometry, both $E$ and the pressure $P$ depend on the radial coordinate $r$ only, and the velocity field  $v=v_P + v_E$ and charge current 
$j=j_P + j_E$ point in radial direction. Integrating 
over the vertical variable $z$ we obtain the fluxes 
of volume and charge, 
\begin{align}
    J_V = \int_{-h/2}^{h/2} dz (v_P + v_E) \equiv - L_{vv} \nabla P + L_{vc} E,
    \label{eq:4}
\end{align}
\begin{align}
  J_C = \int_{-h/2}^{h/2} dz (j_P + j_E) \equiv - L_{cv} \nabla P + L_{cc}  E,
\label{eq11s1}
\end{align}
where the second identity defines the linear transport coefficients $L_{ij}$ with respect to the generalized forces $-\nabla P=-dP/dr$ and $eE$.  

The first term in eq. \eqref{eq:4} arises from the pressure driven flow profile $v_P(z)$. Assuming no-slip boundary 
conditions $v_P(\pm h/2)=0$, the Stokes equation \eqref{eq:9} with $E=0$
is readily integrated, 
\begin{align}
  v_P = -\frac{h^2 - 4z^2}{8\eta}\nabla P,
\end{align}
resulting in 
\begin{align}
 L_{vv} = \frac{h^3}{12\eta}. 
\end{align}
The second term in \eqref{eq:4} accounts for the electro-osmotic velocity profile \cite{Anderson1989}
\begin{eqnarray}
    v_E(z) &=& - \frac{1}{\eta} \int_z^{h/2} dz' \int_0^{z'} dz'' \rho(z'') E
                         \nonumber \\
            &=&  \frac{\varepsilon}{\eta} \left(\psi(z) - \zeta\right) E         ,
\end{eqnarray}
where the second identity follows from twice integrating Gauss' law $\varepsilon\partial_z^2\psi=-\rho$. This leads to the electro-osmotic transport coefficient 
\begin{align}
    L_{vc} = \frac{1}{E} \int_{-h/2}^{h/2}dz v_E(z).
\end{align}

The electric current \eqref{eq11s1} consists of advection of counterions in the Poiseuille flow profile $v_P$,
\begin{align}
  L_{cv} =  \frac{1}{\eta}\int_{-h/2}^{h/2} dz \rho(z)\frac{h^2 - 4z^2}{8},
\label{eq:8}
\end{align}
and Ohm's law with the conductivity  $L_{cc}$. This latter coefficient reads as
\begin{align}
  L_{cc} =  \int_{-h/2}^{h/2} dz \left(\rho 
   \frac{\varepsilon}{\eta} \left(\psi - \zeta \right) 
        +  e^2 (\mu_+ n_+ + \mu_- n_-)\right)
\label{eq:10}
\end{align}
where the first term accounts for advection by the electro-osmotic velocity field 
$v_E$, and the second one for electrophoresis of salt ions, with mobilities $\mu_\pm$.

Electrokinetic phenomena in a channel between two electrolyte 
reservoirs at different electrochemical potential, are 
characterized by a constant streaming current $J_C\neq0$ 
\cite{Gross1968,Alexander1987,Marbach2019}. Contrary to this 
open geometry, the periodically driven squeezing motion of 
Fig. \ref{fig:1} does not allow for a steady current but 
gives rise to the electric field $E$. Strictly speaking, there 
is a small current which develops the space charges related to 
the electric field, $\delta\rho=\varepsilon\mathrm{div}\cdot E$, 
and which vanishes when averaged over one cycle. Because of the 
strong electric interactions, the space charges develop almost 
instantaneously such that the electric field is in phase with 
the pressure gradient, and that advection and conduction currents 
cancel each other in \eqref{eq11s1},
\begin{equation}
J_C =0.
\label{eq:15}
\end{equation}
This relation holds true as long as the charge relaxation time $\tau$ is much shorter than the period of the external driven, $\omega\tau\ll1$.

\subsection{Drag force}

With the above condition of zero charge current, eq. \eqref{eq11s1} 
implies a relation between the radial electric field and the 
pressure gradient,
\begin{align}
  E =  \frac{L_{cv}}{L_{cc}} \nabla P .
  \label{eq:17}
\end{align}
Inserting this in the volume current \eqref{eq:4} and solving 
for the pressure gradient, we find
\begin{align}
\nabla P = -\frac{6\eta rV}{h^3}\frac{1}{1-\xi} ,
\label{eq:14}
\end{align}
where the coupling of the double layer to the flow is accounted for 
by the ratio of off-diagonal and diagonal transport coefficients $L_{ij}$,
\begin{align}
\xi=\frac{L_{vc}L_{cv}}{L_{vv}L_{cc}}.
\label{eq37sc}
\end{align}
From \eqref{eq:14} it is clear that the dimensionless parameter 
$\xi$ describes the effect of charge-flow coupling on the 
hydrodynamic pressure. For $\xi=0$ one recovers the well-known 
expression for the pressure gradient in capillary. The stability 
of the dynamic equations \eqref{eq:4} and \eqref{eq11s1} requires a positive determinant of the matrix of the transport coefficients $L_{ij}$, that is $\det{\mathbf{L}>0}$ or $\xi<1$. 

When integrating the excess hydrodynamic pressure in the capillary,
it turns out convenient to use the variable $h$ instead of $r$. In 
lubrication approximation \eqref{eq:19} one has $dh=dr r/R$ and  
\begin{align}
  P(h) = 6\eta R V \int_{h}^{\infty}  \frac{dh'}{h'^3}\frac{1}{1-\xi(h')}.  
  \label{eq:24}
\end{align}
Finally, the viscous force on the cantilever is given by the 
surface integral of the pressure. With $dS = 2\pi drr =2\pi Rdh$ 
one finds for the drag coefficient
\begin{align}
  F(h_0) = -2\pi R \int_{h_0}^\infty dh P(h).  
  \label{eq:25}
\end{align}
In eq.\eqref{eq:0} we have defined the electroviscous drag 
coefficient through $F=-\gamma V$; the above relations give
\begin{align}
  \gamma = 12 \pi\eta R^2 \int_{h_0}^\infty dh 
  \int_{h}^\infty  \frac{dh'}{h'^3}\frac{1}{1-\xi(h')}. 
  \label{eq:26}
\end{align}

In the absence of electro-viscous coupling, one readily obtains the pressure 
\begin{align}
P_0(h) = \frac{3\eta V R}{h^2}, \;\;\;\; (\xi=0) , 
\end{align}
which is maximum at the centre of the film and vanishes 
as $P_0\propto r^{-4}$ at large radial distance. The corresponding lubrication drag coefficient \cite{Brenner1961},
\begin{align}
  \gamma_0 = \frac{6\pi \eta  R^2}{h_0},      \;\;\;\; (\xi=0) , 
  \label{eq:5}
\end{align}
is by a factor $R/h_0$ larger than the Stokes drag coefficient $6\pi \eta  R V$ on a sphere of radius $R$ in a bulk liquid.

\section{Electroviscous drag coefficient}

\blue{
As a main formal  result of this paper,  eq. \eqref{eq:26} expresses the electroviscous drag enhancement in terms of the coupling coefficient  $\xi$ which quantifies the charge-flow coupling. Here, we evaluate eq. \eqref{eq:26} both analytically and numerically. }

\subsection{Wide-channel approximation $h\gg\lambda$}

If the height of the water film is much larger than the 
Debye length, the electrostatic potential 
is given by \cite{Andelman2006}
\begin{align}
    \psi = -\frac{4k_BT}{e}\mathrm{arctanh}(\beta e^{-z/\lambda}),
    \label{eq:13}
\end{align}
where the parameter
\begin{equation}
    \beta = \frac{\sqrt{1+(2\pi\ell_B\lambda\sigma)^2} - 1}
       {2\pi\ell_B\lambda\sigma}
 \label{eq:13a}
\end{equation}
depends on the Debye length $\lambda$ and the surface charge density $\sigma$. 

In this case, there are analytical expressions for the transport 
coefficients $L_{ij}$. The off-diagonal terms are given by the 
Helmholtz-Smoluchowski electrophoretic mobility,
\begin{align}
 L_{vc} =   -\frac{h\varepsilon\zeta}{e\eta} = -\frac{h\hat{\zeta}}{4\pi \eta l_B},
\end{align}
where in the second identity we define the dimensionless zeta 
potential in units of the thermal energy $\hat \zeta=e\zeta/k_BT$. 
{}{
The electrical conductivity reads as
\be
 L_{cc} = \frac{\sinh(\hat{\zeta}/4)^2}{\pi^2\eta\lambda\ell_B^2} 
                   + \sum_\pm \mu_\pm n_0 \( h - \frac{4\beta\lambda}{\beta\mp1} \) ,
\ee
where the first term accounts for electro-osmotic advection and the second for ion electrophoresis, with surface contributions parameterized by $\beta$. 
}

{}{
For wide channels, $h\gg\lambda$, the conductivity is dominated by bulk ion 
electrophoresis. Discarding the electro-osmotic and surface terms, and using the definition 
of the screening length \eqref{eq:1b}, results in the coupling parameter 
\begin{align}
\xi = \frac{\lambda_*^2}{2h^2},
\label{eq:7}
\end{align}
with the length scale 
\begin{align}
  \lambda_* = 6\hat{\zeta}\sqrt{\frac{a}{\ell_B}}\lambda. 
  \label{eq:21}
\end{align}
Here and in the following, the mobilities are expressed through ion radii, $\mu_\pm=1/6\pi\eta a_\pm$, with the mean value $1/a=2/a_++2/a_-$. 
}

\begin{figure}[tb]
		\includegraphics[width=\columnwidth]{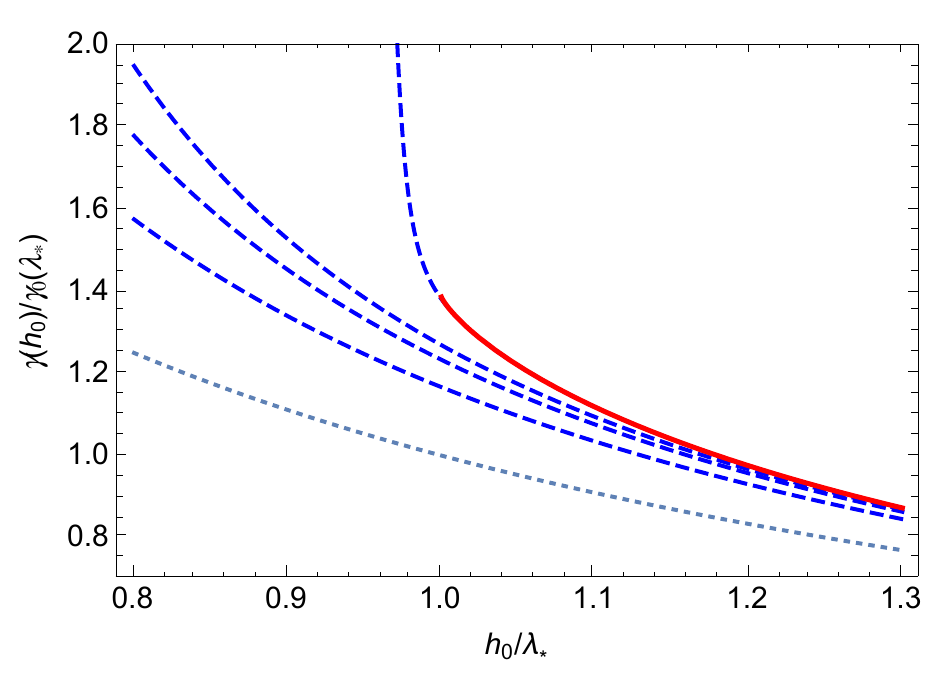}
		\caption{Electroviscous enhancement of the drag coefficient $\gamma(h_0)$, in units of the purely viscous coefficient $\gamma_0$ at $h_0=\lambda_*$. In the absence of charge-flow coupling as in (\ref{eq:5}), the dotted line gives $\gamma_0(h_0)/\gamma_0(\lambda_*)=\lambda_*/h_0$. Dashed lines are calculated from the perturbation series \eqref{eq:152} for $\gamma$, truncated	at $(\lambda_*/h_0)^{2n}$ with $n=1,2,3,100$. The full line represents the complete series \eqref{eq150sc}, which is defined for $h>\lambda_*$ only.}
	\label{fig:2}
\end{figure}

Then the pressure \eqref{eq:24} and the drag coefficient \eqref{eq:26}  
can be integrated in closed form,
\begin{align}
      \frac{\gamma}{\gamma_0} = \frac{h_0}{\lambda_*}\ln\frac{h_0+\lambda_*}{h_0-\lambda_*}+\frac{h_0^2}{\lambda_*^2}\ln\frac{h_0^2 - \lambda_*^2}{h_0^2}.
	\label{eq150sc}
\end{align}
In Fig. \ref{fig:2} we plot $\gamma$ as a (red) solid line. At the distance $h_0=\lambda_*$ the electroviscous coupling parameter $\xi$ is equal to unity and, as a consequence, a logarithmic branch point appears in the pressure integral \eqref{eq:24}, resulting in $\gamma/\gamma_0=2\ln2\approx1.39$. At smaller distances the wide-channel approximation for pressure and force integrals is not defined.

It turns out instructive to rewrite \eqref{eq150sc} as a series 
in powers of $\lambda_*/h_0$, 
\begin{align}
      \frac{\gamma}{\gamma_0} =  1 + \frac{1}{6} \frac{\lambda_*^2}{h_0^2} + \frac{1}{15} \frac{\lambda_*^4}{h_0^4} + 
      \frac{1}{28} \frac{\lambda_*^6}{h_0^6} + ... 
	\label{eq:152}
\end{align}
In Fig. \ref{fig:2} we plot this series truncated at $(\lambda_*/h_0)^{2n}$ with $n=1,2,3,100$, and compare both with \eqref{eq150sc}  and with the uncoupled lubrication drag coefficient \eqref{eq:5}. Retaining a few terms only, suggests a smooth behavior, whereas Eq. \eqref{eq150sc} is defined for $h_0\ge \lambda_*$ only. The first correction term, proportional to $\lambda_*^2/h_0^2$, corresponds to the electroviscous coefficient of Bike and Prieve \cite{Bike1990}. 

Noting that the ion radius is usually smaller than the Bjerrum 
length $\ell_B=0.7$ nm, and $\hat\zeta$ of the order of unity, 
one finds that $\lambda_*/\lambda$ takes values between 1 and 10. 

\begin{figure}[tb]
	\includegraphics[width=\columnwidth]{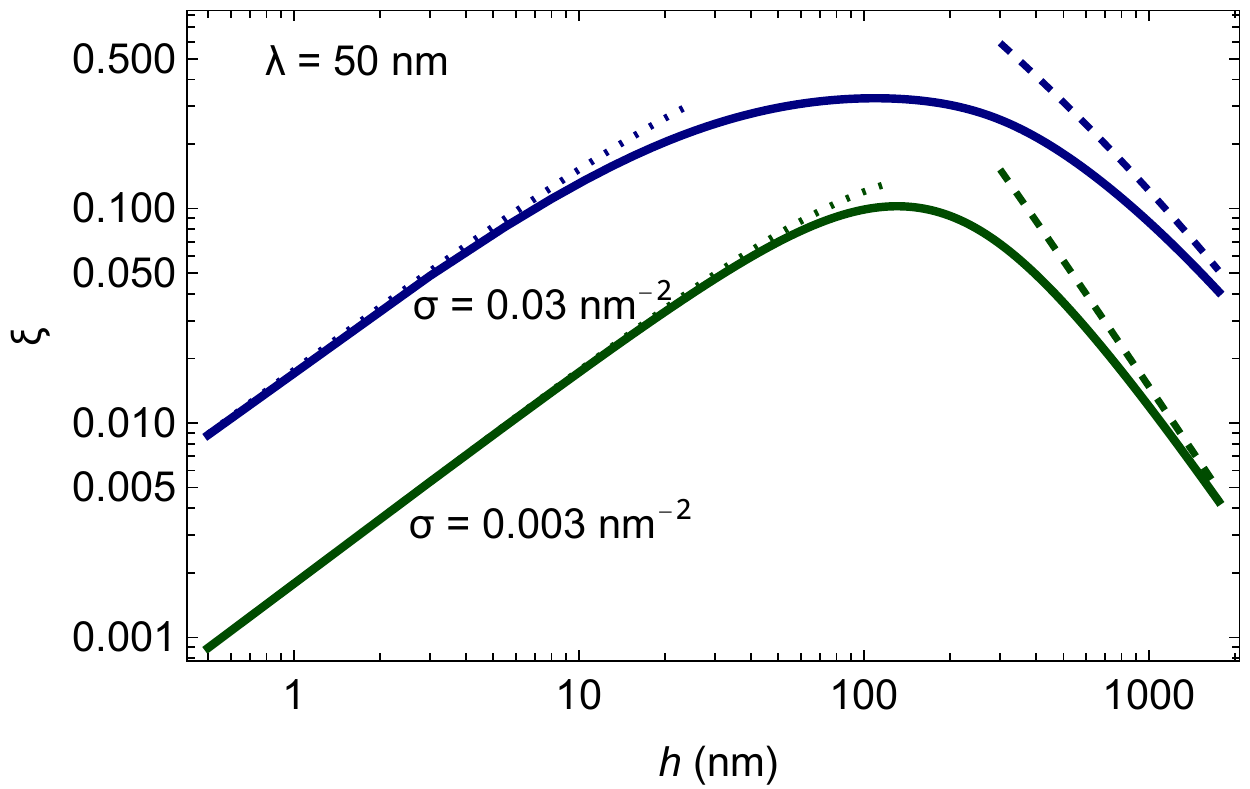}
	\caption{Numerical calculation of the electroviscous 
	coupling parameter $\xi$ as a function of $h$, for 
	surface charge density $\sigma=0.003$ and $0.03\,\mathrm{nm}^{-2}$, 
	and Debye length $\lambda=50\,\mathrm{nm}$. 
	Dotted and dashed lines correspond to the approximations 
	of narrow and wide channels, respectively, whereas the 
	solid lines give the numerical solution.}
	\label{fig:xi}
\end{figure}

\subsection{Narrow-channel approximation}

In the case of a narrow channel, $h\ll\lambda$, the overlapping 
double layers of the surfaces result in a constant charge density
\begin{align}
    \rho = \varepsilon\partial_z^2 \psi = 2\sigma/h, 
\end{align}
in other words, the counterions form a homogeneous gas \cite{Andelman2006}. 
The electrostatic potential is readily integrated,
\begin{align}
    \psi(z) = \frac{k_BT}{e}\left(\ln m - \frac{4\pi\ell_B\sigma}{h} z^2 \right) ,
    \label{eq:32}
\end{align}
where the  parameter $m$ describes the finite value of the potential $\psi(0) = (k_BT/e) \ln m$ at $z=0$. 

{}{
With these expressions for $\rho$ and $\psi$ the transport 
coefficients are readily calculated. Retaining contributions 
of leading order in $h$ only, we find
\begin{equation}
 L_{vc} =  \frac{e \sigma h^2}{6\eta},  
     \;\; L_{cc} = \frac{e^2 \sigma}{3\eta a_+}, 
\end{equation}
resulting in the coupling parameter 
\begin{align}
  \xi = \sigma a_+ h .
\label{eq:37}
\end{align}
Note that, for narrow channels, the conductivity is independent of salinity and gap height \cite{stein2004surface}, whereas the parameter $\xi$ is linear in  $h$. 
}

\subsection{Numerical evaluation of $\xi$ and $\gamma$} 

In the general case, the electrostatic potential is obtained in 
terms of the Jacobi elliptic function $\cd(u|m^2)$ \cite{Abramowitz1964}, 
\begin{align}
\psi(z) = \frac{k_BT}{e} \left[ \ln m + 2\ln\cd\left(\left.\frac{z}{2\lambda\sqrt{m}}\right|m^2 \right)\right].
\label{eq:48}
\end{align}
Because of $\cd(0|m^2)=1$, the second term vanishes at $z=0$, and 
the potential at $z=0$ is determined by $\ln m$. The parameter $m$ 
depends on the ratio of the channel height and the Debye length: For 
$h\gg\lambda$ one has $m=1$ and recovers the analytic 
expression \eqref{eq:13} for a charged surface limiting an 
infinite half-space. In the narrow-channel limit one finds 
  \begin{align}
      m = \frac{h n_0}{2x \sigma},
        \;\;\;\; (h n_0\ll\sigma) ,
  \label{eq:49}  
  \end{align}
and expanding the Jacobi function to second order in $z$, one 
recovers the potential defined in eq. \eqref{eq:32} above.

\begin{figure}[t]
	\includegraphics[width=\columnwidth]{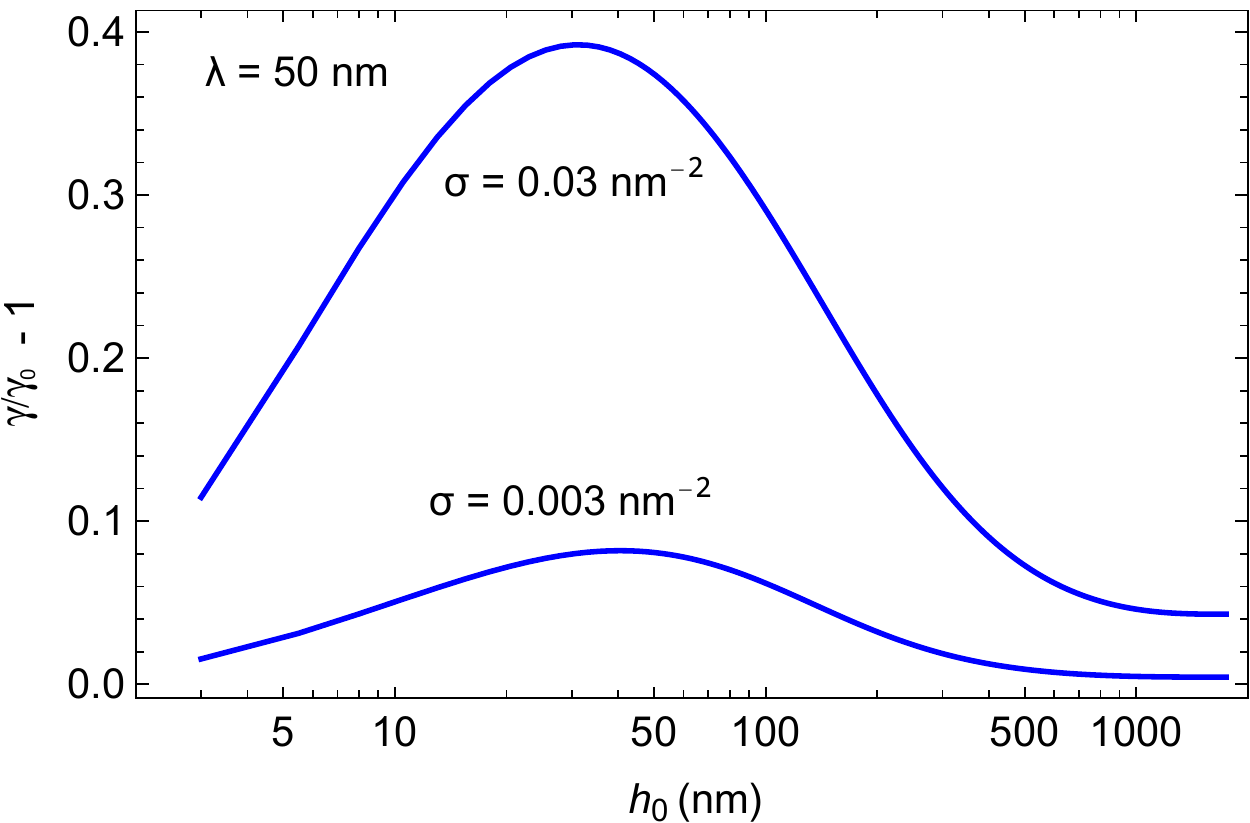}
	\caption{Electroviscous drag enhancement $\gamma/\gamma_0-1$ as a function of $h_0$, for different values of the surface charge density $\sigma$.}
	\label{fig:forcestatic}
\end{figure}

{}{
The electric potential is calculated numerically from \eqref{eq:48} 
with the boundary condition \eqref{eq:2}. Then the electroviscous 
coupling parameter $\xi$ defined in \eqref{eq37sc}, is obtained by 
performing the integrals \eqref{eq:8} and \eqref{eq:10} for a 
given film distance $h$. The numerical results are given in terms of the gap height $h_0$, the surface charge density $\sigma$, and the Debye screening length $\lambda$. We use the viscosity of water at room temperature,  $\eta=0.9\times 10^{-3}$ Pa.s, and the ion mobilities $\mu_\pm = 1/6\pi\eta a_\pm$ with the radii of sodium $a_+=1.9\,\mathrm{\mathring{A}}$ and of chlorine  $a_-=1.3\,\mathrm{\mathring{A}}$ \cite{Bulavin2011}.
}
 
Fig. \ref{fig:xi} shows the variation of $\xi$  as a function of 
$h$ for different values of surface charge concentration $\sigma$, 
and compares with narrow-channel and 
wide-channel approximations. As a surprising feature, $\xi$ is 
roughly linear in $\sigma$. The log-log plot shows the power laws 
$\xi \propto h$ and $\xi \propto h^{-2}$, in the limits of narrow 
and wide channels, respectively. The maximum occurs at 
$h_\mathrm{max}\approx 3\lambda$. The narrow-channel result \eqref{eq:37} 
provides a good description for $h\le\lambda$, whereas the 
wide-channel expression \eqref{eq3s2} converges for 
$h\gg\lambda_*$ only. In the intermediate range, which covers 
at least one decade in $h$, neither of them is valid.

In Fig. \ref{fig:forcestatic} we plot the enhancement factor 
$\gamma/\gamma_0-1$ of the viscous force \eqref{eq:25}, with 
parameters as in Fig. \ref{fig:xi}. As expected for the 
electroviscous coupling parameter $\xi$, there is a maximum 
at $h_0\approx\lambda$. The enhancement factor depends equally 
strongly on the surface charge and the Debye length. 

\section{Charge regulation}\label{section:CR}

So far we have assumed that the surface charge 
density $\sigma$ remains constant upon varying 
the film height $h_0$. This is not the case, 
however, for weakly dissociating acidic groups 
$\mathrm{HA}$ which release and recover protons 
according to \cite{Nin71}
\begin{align}
\mathrm{HA\rightleftharpoons H^+ + A^-}.
\end{align}
For narrow channels the potential \eqref{eq:48} 
takes a finite value $\psi(0)=(k_BT/e)\ln m$ at $z=0$, 
which  favors recombination of the surface groups, 
thus reducing the effective charge density $\sigma$ 
and surface potential $\zeta$.  

A simple and widely studied model relies on the 
dissociation constant 
\begin{align}
  {\cal Z} = \frac{\mathrm{[H^+][A^-]}}{\mathrm{[HA]}}
    = n_s\frac{ \alpha}{1-\alpha},
\end{align}
where we have defined the dissociated fraction 
$\alpha$ and the hydronium concentration at the 
surface $n_s=e^{-e\zeta/k_BT}\mathrm{[H^+]}_\infty$. 
Solving for $\alpha$ one finds the fraction of 
dissociated sites
\begin{align}
  \alpha= \frac{ 1}{1+n_s/{\cal Z}},
\end{align}
and the number density of surface charges,
\begin{align}
  \sigma = \frac{ \alpha}{S}.
\end{align}
The electrostatic potential is obtained by 
closing the above relations with the boundary 
condition \eqref{eq:2}. The area per site $S$ 
is chosen such that at large distance (where 
$\zeta=\zeta_\infty$), $\sigma$ takes the 
value indicated for the case of constant charge. 

\begin{figure}[t]
    \includegraphics[width=\columnwidth]{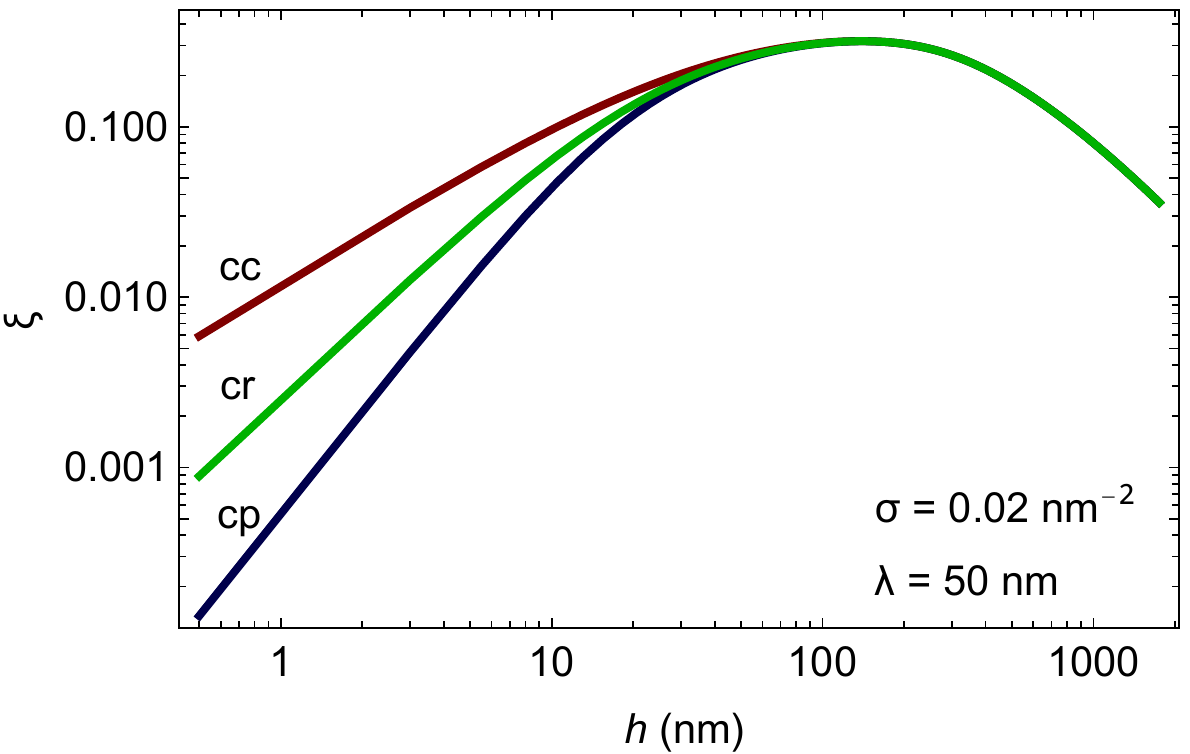}
	\caption{Electroviscous coupling parameter $\xi$ 
	as a function of the distance $h$, for constant 
	charge (cc), charge regulation (cr), and constant 
	potential (cp).}
	\label{fig:xi_CR}
\end{figure}

An alternate approach, which is often used 
for systems with more complex charging 
procedure but essentially leads to the same 
results, is via a proper minimization of 
the relevant thermodynamic potential \cite{Maj18}.

In the following we compare the electrostatic 
and electroviscous properties calculated at 
constant charge (cc) with the charge-regulated 
case (cr), and also with that of constant 
potential (cp), where the boundary condition \eqref{eq:2} 
is replaced with 
\begin{align}
  \psi(\pm h/2) = \zeta_\infty.
\end{align}
Here $\zeta_\infty$ is the surface potential 
at large distance, calculated with the surface 
charge $\sigma$ according to \eqref{eq:13}. 
All curves labelled ``cr'' are calculated with 
${\cal Z}=10^{-3}$ M.

\subsection{Electroviscous coupling}

In Fig. \ref{fig:xi_CR} we plot the coupling 
parameter $\xi$ for the cases of constant 
charge and constant surface potential, and 
observe a behavior similar to what has been 
reported previously for the disjoining 
pressure \cite{Markovich_2016}: At distances 
smaller than the screening length, $h<\lambda$, 
the curves of $\xi$ for different boundary 
conditions diverge significantly. Yet note 
that the electroviscous coupling is strongest 
in the range $\lambda < h < 10 \lambda$, where 
charge regulation is of little importance.

\begin{figure}[tb]
	\includegraphics[width=\columnwidth]{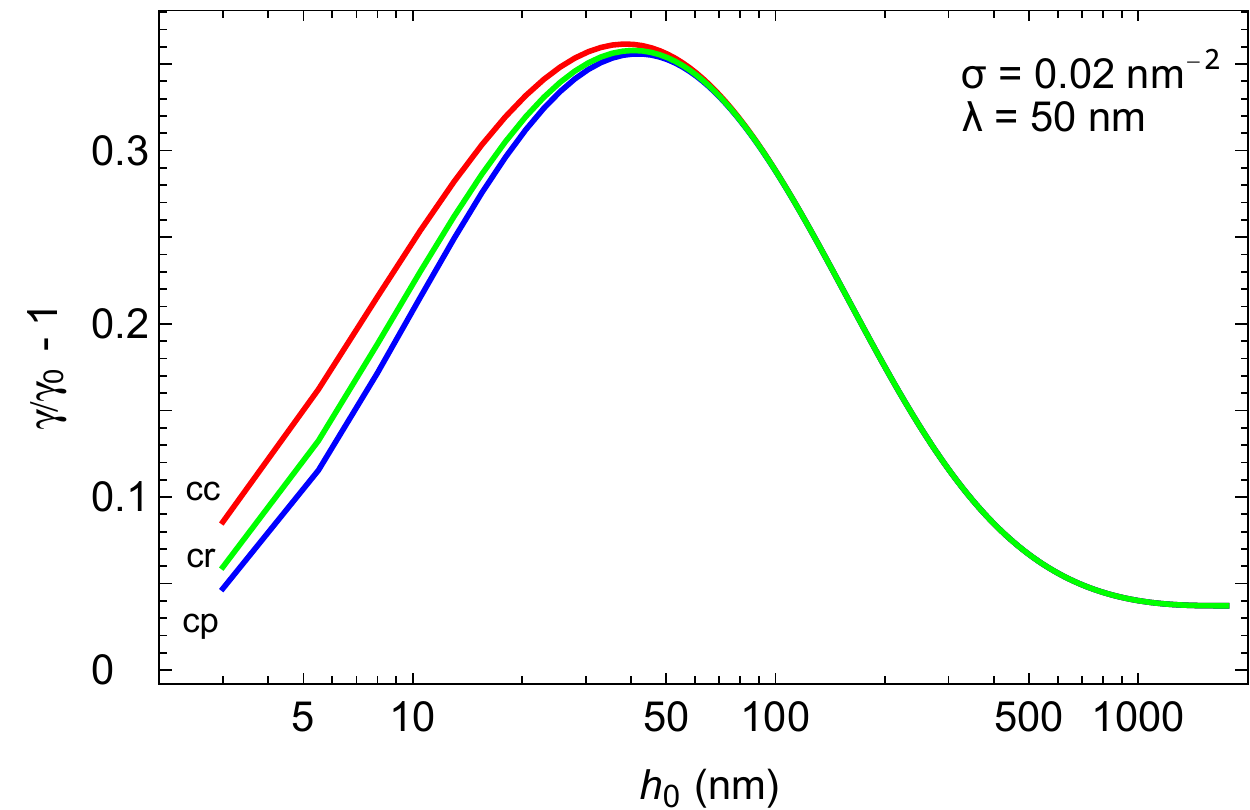}
	\caption{Electroviscous enhancement $\gamma/\gamma_0-1$, as a function of the distance $h_0$ for different boundary conditions.}
	\label{fig:F_CR}
\end{figure}

The electroviscous enhancement of the drag force $\gamma$ 
with respect to the uncoupled expression $\gamma_0$ is 
shown in Fig. \ref{fig:F_CR}. The maximum occurs at a 
distance slightly below the screening length. For the 
given electrostatic parameters, it reaches a value 
of about $35\%$, which depends little on the electrostatic 
boundary condition. The electroviscous drag component 
disappears at much higher distances of about $10\lambda$.

\subsection{Disjoining pressure and static repulsion}

\begin{figure}[t]
	\includegraphics[width=\columnwidth]{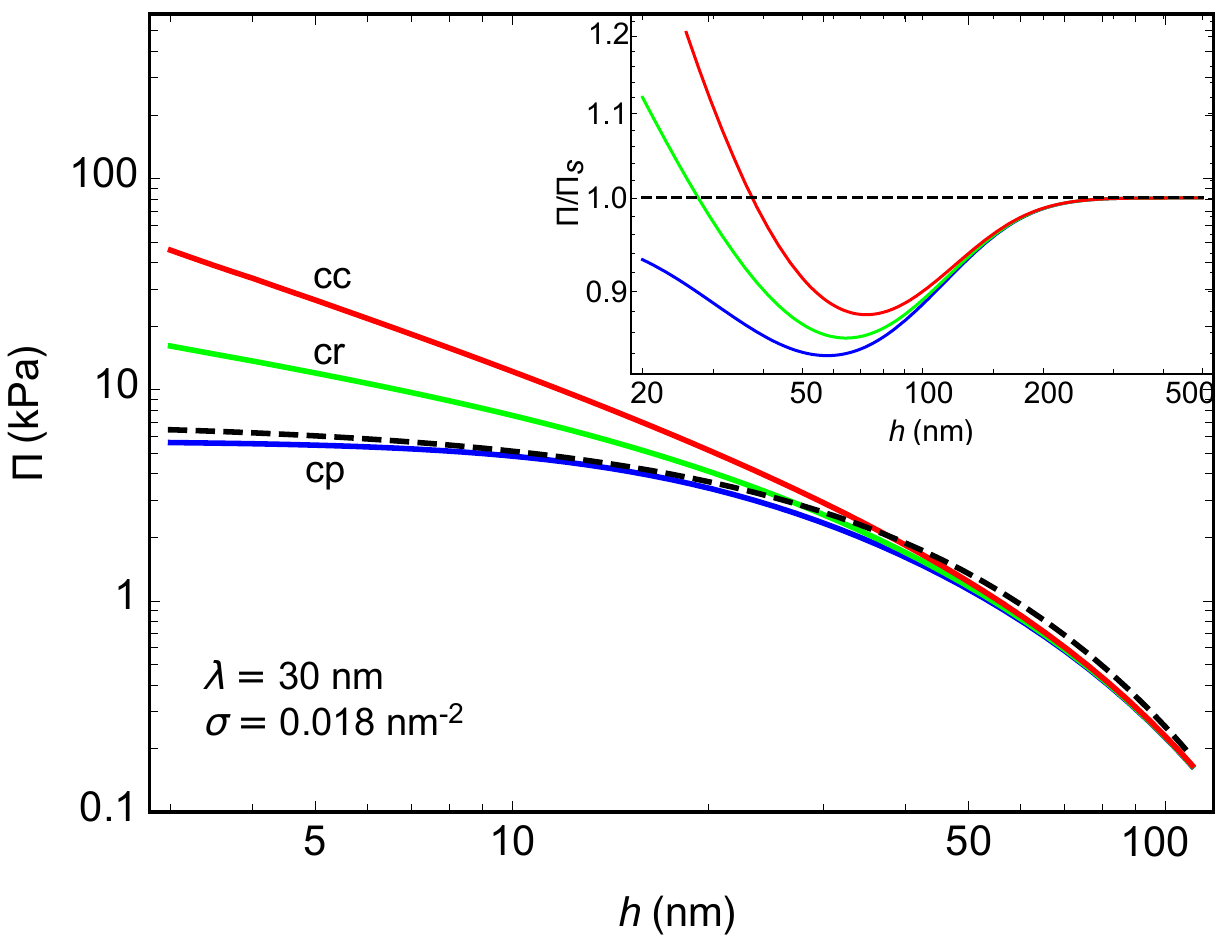}
	\caption{Disjoining pressure between charged 
	surfaces as a function of the distance $h_0$. The 
	solid curves give the numerical solution 
	\eqref{eq:Pi} for constant surface charge 
	$\sigma=0.018 \mathrm{nm^{-2}}$ (cc), constant 
	potential $\zeta$ (cp), and the charge-regulated 
	intermediate case (cr) with dissocation constant 
	${\cal Z}=10^{-3}$M. The approximative expression 
	\eqref{eq:Pi_approx} is plotted as dashed line. 
	The inset shows the ratio $\Pi/\Pi_s$, 
	highlighting the deviation of the disjoining 
	pressure $\Pi$ from the approximate expression 
	$\Pi_s$, which sets in well above 200 nm.}
	\label{fig:Pi}
\end{figure}

Now we consider the static repulsive force arising from the overlap of the diffuse layers on the opposite surfaces, and which is independent of the external driving. According to \eqref{eq:48} the potential at $z=0$ 
reads as $\psi(0) = (k_BT/e)\ln m$, and the disjoining 
pressure \eqref{eq:Pi} is determined by the parameter $m$,
\begin{align}
\Pi=n_0k_BT\left(m+\frac{1}{m}-2\right).
\end{align}
In Fig.	\ref{fig:Pi} we plot $\Pi$ calculated for 
constant charge (cc), constant potential (cp), and 
charge regulation (cr). For distances shorter than 
the screening length, these different boundary 
conditions result in significant differences. In 
agreement with previous work, we find a constant 
pressure for cp \cite{Andelman2006} and power laws 
$\Pi\propto h^s$ with $s=-1$ and $-\frac{1}{2}$ 
for cc and cr, respectively \cite{Markovich_2016}.

The dashed line corresponds to the widely used 
approximation \cite{Israelachvili1991} 
\begin{align}
	\Pi_s(h) = 64 \beta^2 n_0 k_BT e^{-h/\lambda}, 
	\;\;\; (h\gg\lambda),
	\label{eq:Pi_approx}
\end{align}
which relies on the linear superposition of the 
double layers at the opposite surfaces, and where 
the parameter $\beta = \tanh(e\zeta_\infty/4k_BT)$
is given by the surface potential $\zeta_\infty$ 
at $h_0\rightarrow\infty$, as defined in eq. \eqref{eq:13a}.

The repulsive force \eqref{eq:60} between the two 
surfaces is calculated in Derjaguin approximation, 
in analogy to \eqref{eq:25}, resulting in
\begin{align}
 K = 2\pi R \int_{h_0}^{\infty} dh \Pi(h). 
 \label{eq:60a}
\end{align}
For the pressure in superposition approximation we 
obtain $K_s = 2\pi R\lambda \Pi_s(h_0)$ and, after 
expressing the salt content through the Debye length,
\begin{align}
	K_s = \frac{16 R \beta^2 k_BT}{\lambda \ell_B} e^{-h_0/\lambda},
	\;\;\;\;\; (h_0\gg\lambda).
	\label{eq:electrostatic force}
\end{align}

\begin{figure}[tb]
	\includegraphics[width=\columnwidth]{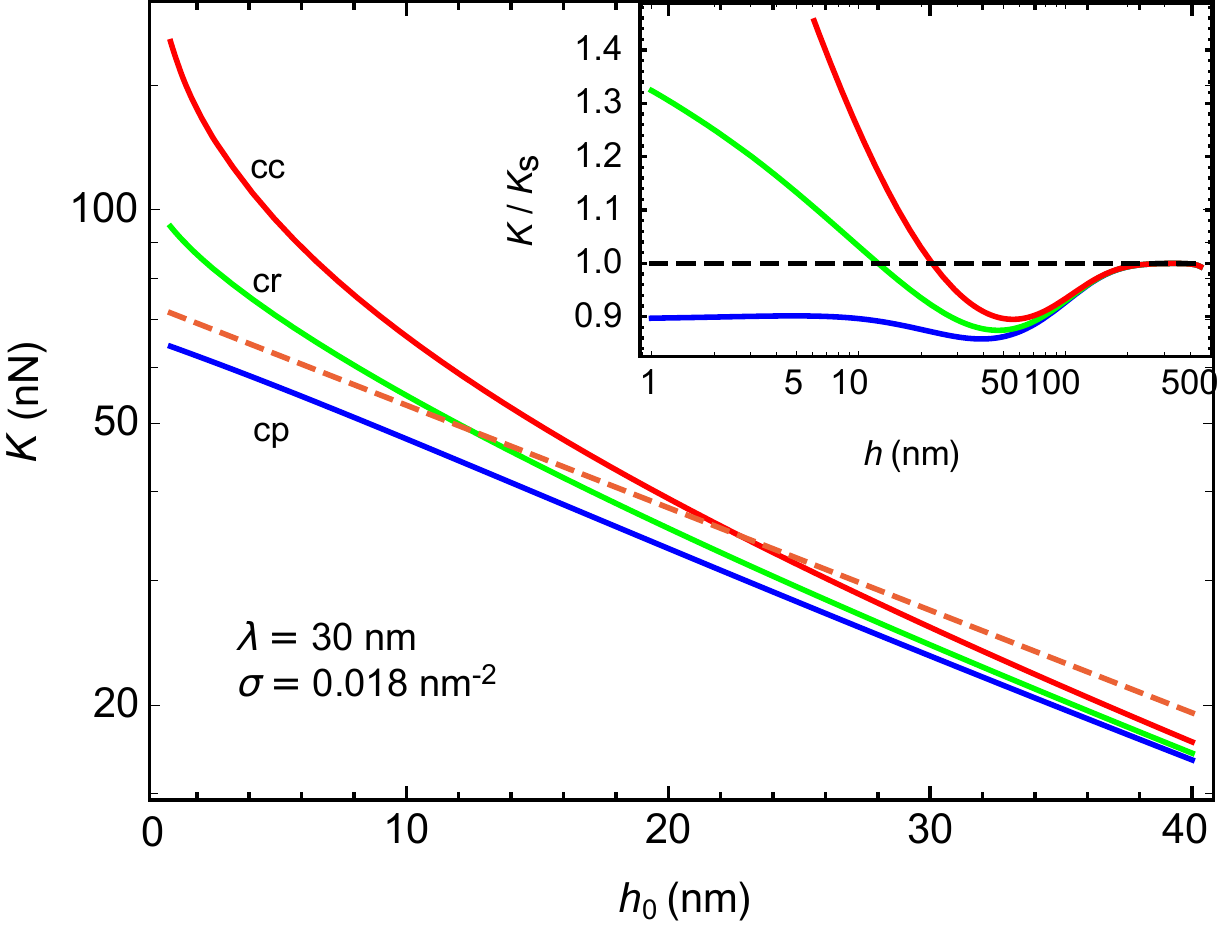}
	\caption{Static force between charged surfaces as a function of the distance $h_0$. The solid curve give the numerical solution \eqref{eq:60} for constant charge (upper red), constant potential (lower blue), and charge regulation (middle green). The approximative expression \eqref{eq:electrostatic force} is plotted as dashed line. The inset shows the ratio $K/K_s$; note that all curves coincide at large distance, which is not visible in the main figure.}
	\label{fig:staticforce_lin}
\end{figure}

A comparison of the numerically exact force $K$ 
with the exponential approximation $K_s$ is given 
in Fig. \ref{fig:staticforce_lin}. Both expressions 
agree beyond 200 nm, or $h_0 >7\lambda$. The inset shows 
that the force calculated for constant potential (cp) 
remains about 10$\%$ below $K_s$, whereas those for 
constant or regulated charge (cc or cr) show a more 
complex behavior: they first decrease below $K_s$ yet 
at even smaller $h_0$ by far exceed the analytic 
approximation $K_s$ \cite{Israelachvili1991}.

\section{AFM force measurement}
\subsection{Experimental detail}

We performed a dynamic AFM measurement with colloidal 
probe following the method described in \cite{maali2013precise}. 
A spherical borosilicate particle (MO-Sci Corporation) 
with a radius of $R=\mathrm{47 \pm 1~\mu m}$ was 
glued at the end of a cantilever (CSG30, NT-MDT) using 
epoxy (Araldite, Bostik, Coubert). The stiffness of 
the ensemble of cantilever and particle was 
 calibrated by the drainage method~\cite{Craig2001}, resulting in $k_\mathrm{c}=0.8\pm0.1\mathrm{N/m}$. The resonance 
frequency and bulk quality factor were obtained from 
the thermal spectrum as $\omega_0/2\pi=3340$ Hz and 
$Q=4.7$, respectively. 

\begin{figure}[t]
	\includegraphics[width=\columnwidth]{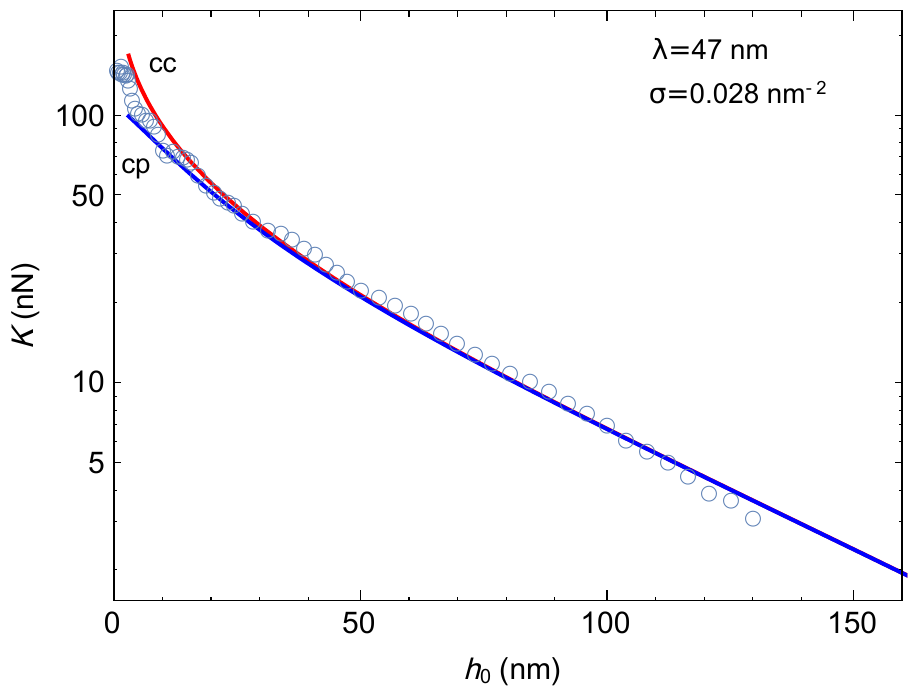}
	\caption{Static repulsion $K$ between the AFM sphere and the solid surface, as a function of the distance $h_0$. The squares give experimental data. The blue and red curves are calculated from \eqref{eq:60a} for constant potential and constant surface charge, respectively, with the parameter values $R=55\mathrm{\mu m}$, surface charge density $\sigma=0.028\,\mathrm{nm}^{-2}$ and  screening length $\lambda=47\,\mathrm{nm}$.} 
	\label{fig:static_force}
\end{figure}

The experiment was performed using an AFM (Bioscope, Bruker, USA) equipped with a liquid cell (DTFML-DD-HE) which allows us to work in liquid environment. The mica surface was driven by a piezo (Nano T225, MCL Inc., USA) to approach the particle with a very small velocity such that the drainage force can be neglected, and meanwhile the probe was also driven with a base oscillation amplitude $A_\mathrm{b}=3.5$ nm  and frequency of 
$\omega/2\pi = 100$ Hz. The amplitude $A$ and phase 
$\varphi$ of the cantilever deflection were measured 
by a Lock-in Amplifier (Signal recovery, Model:7280), 
and the DC component of the cantilever deflection was 
also recorded, from which the separation distance 
$h_0$ and electrostatic force $K$ between 
the sphere and the mica surface were extracted. The 
mica surface and cantilever probe are immersed in 
low-salinity water. We also performed control 
experiments at large salinity. All measurements were done at room temperature $21^{\circ}$C.

\subsection{Static force}

Fig. \ref{fig:static_force} shows the electrostatic repulsive force between the mica surface and the colloidal probe. The data roughly show an exponential behavior, as expected for a screened double-layer interaction. The upper (red) curve  is calculated from eq. \eqref{eq:60a} for constant charge number density $\sigma=0.028\,\mathrm{nm^{-2}}$, and the lower (blue) one for constant surface potential $\zeta=-95\,\mathrm{mV}$. In the range where both curves coincide, $h_0>\lambda$, the best fit is obtained with a screening length $\lambda=47\,\mathrm{nm}$, corresponding to an electrolyte strength $n_0=43\,\mathrm{\mu M}$.

\begin{figure}[t]
\includegraphics[width=\columnwidth]{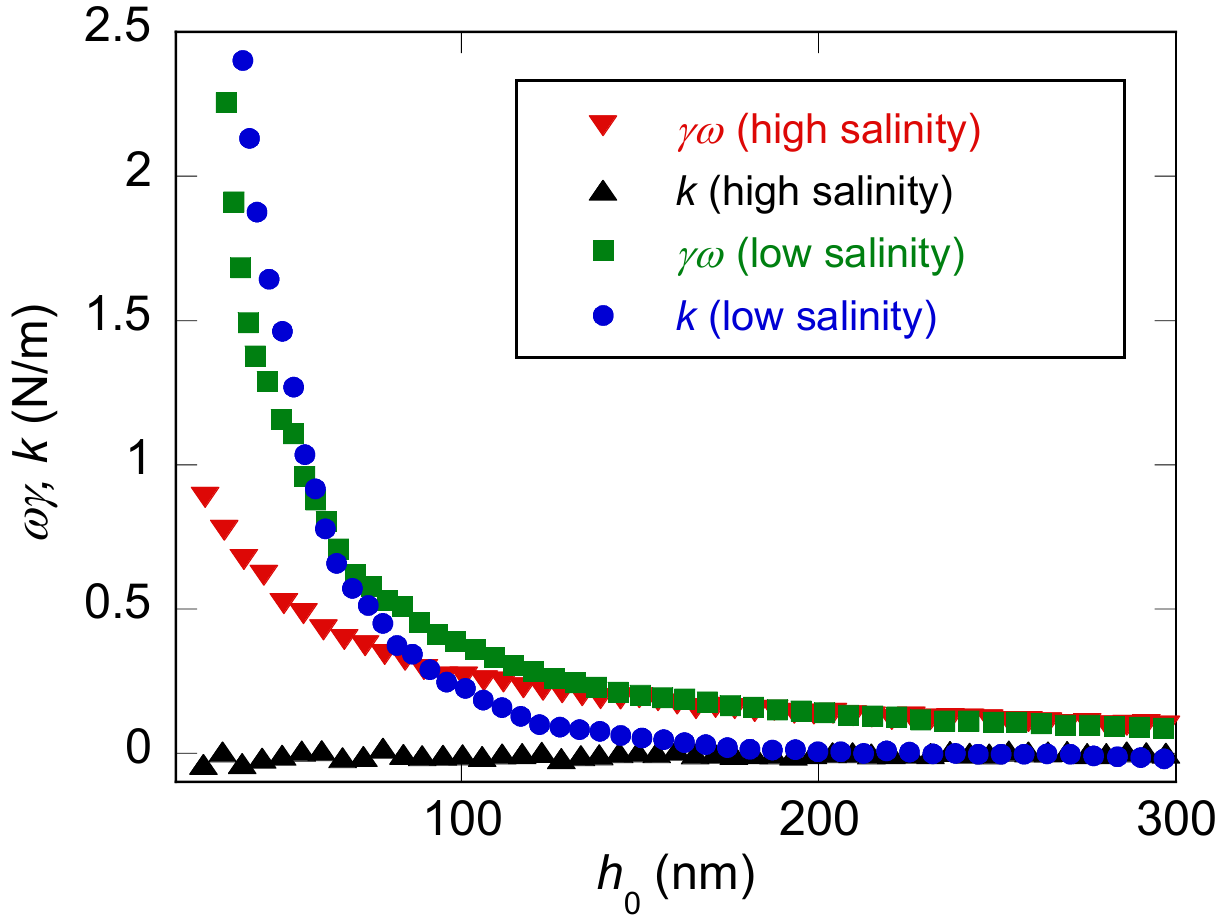}
	\caption{Real and imaginary parts $k$ and $\omega\gamma$ of the response function, measured at a vibrational frequency $\omega/2\pi=100$ Hz and  at low or high salinity, as a function of the distance $h_0$.}
\label{fig:viscoelastic_force}
\end{figure}

\subsection{Spring constant and drag coefficient}

Driving of the probe induces an oscillation of the tip-surface distance according to $h_0+Z(t)$. Modelling the cantilever as a damped harmonic oscillator \cite{Liu2018} and solving its equation of motion for the force $F$ exerted by the surrounding liquid, we obtain in complex notation
\begin{align}
    F=-k_\mathrm{c}Z\left(1-\left(\frac{\omega}{\omega_0}\right)^2+i\frac{\omega}{\omega_0 Q} \right)\frac{A\mathrm e^{i\varphi}-A_{\infty}\mathrm e^{i\varphi_{\infty}}}{A\mathrm e^{i\varphi}+A_\mathrm{b}},
    \label{eq:69}
\end{align}
with amplitude $A$ and phase $\varphi$ of the mica surface. The tip-surface distance reads as $Z(t)=e^{i\omega t}(Ae^{i\varphi} +A_b)$, and the values $A_{\infty}$ and $\varphi_{\infty}$ are measured far from the surface, where the viscoelastic force $F$ is negligible. All measurements are done in the linear-response regime $|Z|\ll h_0$. 

{}{
In view of eq. \eqref{eq:0} we split $F/Z$ in its real and imaginary components. Writing the velocity as $V=i\omega Z$, we readily obtain the complex response function, 
\begin{equation}
    F=-(k+i\omega\gamma)Z,
    \label{eq:70}
\end{equation}
where the ``spring constant'' $k$ and the drag coefficient $\gamma$ account for the  elastic and viscous components of the tip-surface interactions.
}

{}{
In Fig. \ref{fig:viscoelastic_force} we plot the measured real and imaginary coefficients as a function
of the separation distance $h_0$ at low or high salinity, at the oscillation frequency of $\omega/2\pi=100$ Hz. 
At large salinity electrokinetic effects disappear because of electrostatic screening, and  $k$ vanishes accordingly, whereas the drag coefficient follows the law $\gamma_0\propto 1/h_0$, expected from Stokes hydrodynamics \cite{Happel1963}. Quite a different behavior occurs iat low salinity, where we observe a strong elastic component $k$ which decays roughly exponentially with $h_0$, and an electroviscous enhancement  of the drag coefficient. 
}

\begin{figure}[t]
	\includegraphics[width=\columnwidth]{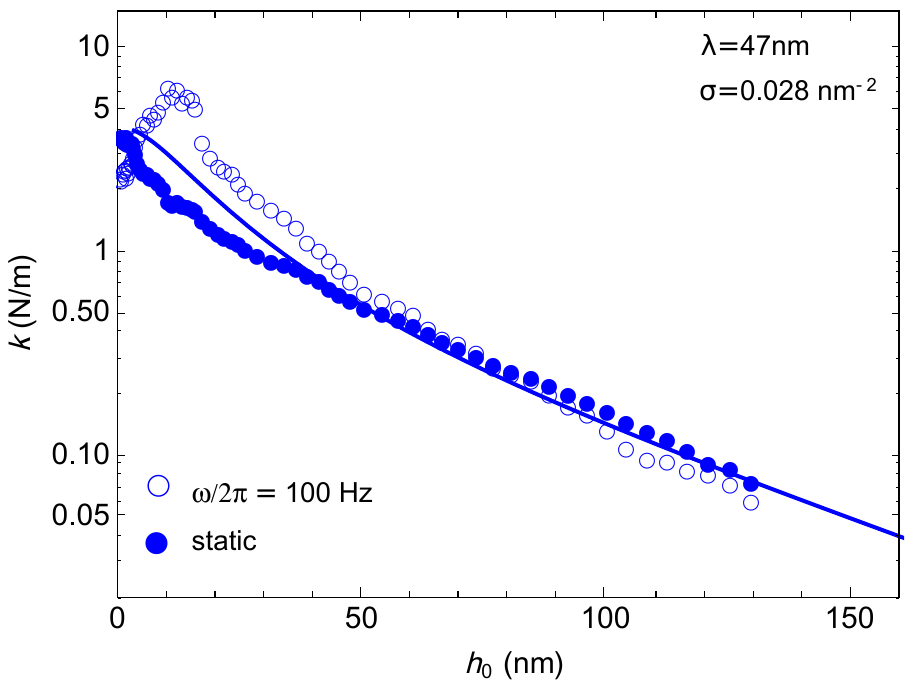}
		\caption{Elastic response $k$ measured at $\omega/2\pi=100$ Hz (open symbols) and for the static case (full symbols), as a function of $h_0$. The solid line give the static rigidity $k$, calculated from eq. \eqref{eq:61} for constant potential. The experimental data are binned, such that each point corresponds to the mean value of 100 measured values.}
		\label{fig:k}
\end{figure}

{}{
In Figs. \ref{fig:k} and \ref{fig:F_exp}, the experimental findings are compared with theory. Regarding the elastic response, Fig. \ref{fig:k}  shows both the static stiffness $-dK/dh_0$ (full symbols) and the dynamic response $k(\omega)$ at finite frequency $\omega/2\pi=100$ Hz (open symbols). The theory curve represents the spring constant \eqref{eq:61}, which is related to the variation of the disjoining pressure with distance and which is calculated from \eqref{eq:Pi} at constant potential (cp). The data roughly follow the exponential law expected for double-layer interactions, and they  provide strong evidence that the dynamic elastic response $k(\omega)$  comprises a frequency dependent contribution which is most significant at small distances, $h_0<\lambda$, and which is not captured by the electrostatic disjoining pressure $\Pi$. 
}

{}{
In Fig. \ref{fig:F_exp} we plot the viscous response function $\omega\gamma$. At high salinity, the electric double-layer is thin ($\lambda<1$ nm), such that charge-flow coupling effects are absent. Indeed, the drag coefficient is well fitted by the viscous contribution $\gamma_0=6\pi \eta R^2 /h_0$, as expected from \eqref{eq:5}. At low salinity, the large screening length $\lambda=47$ nm, comparable to $h_0$, results in charge-advection and electro-osmotic flow, which increase the hydrodynamic pressure and thus enhance the drag coefficient. The theory curve is calculated numerically from eq. \eqref{eq:26}, with the same parameters $\sigma=0.028 \mathrm{nm}^2$ and $\lambda=47$ nm as in Figs. 
\ref{fig:static_force} and \ref{fig:k}. If the overall behavior of the data is rather well described by the theoretical expression, a significant discrepancy occurs for small gaps, where the data exceed the theoretical curve by up to 60$\%$. Comparison with the elastic coefficient shown in Fig. \ref{fig:k},  suggests a frequency dependence of the dynamic response function $k(\omega)+i\omega\gamma(\omega)$, which is not captured by the quasistatic coefficients $k$ and $\gamma$ derived in the present work.}

\section{Discussion} 

\subsection{Validity of the wide channel approximation}
 
\begin{figure}[t]
	\includegraphics[width=\columnwidth]{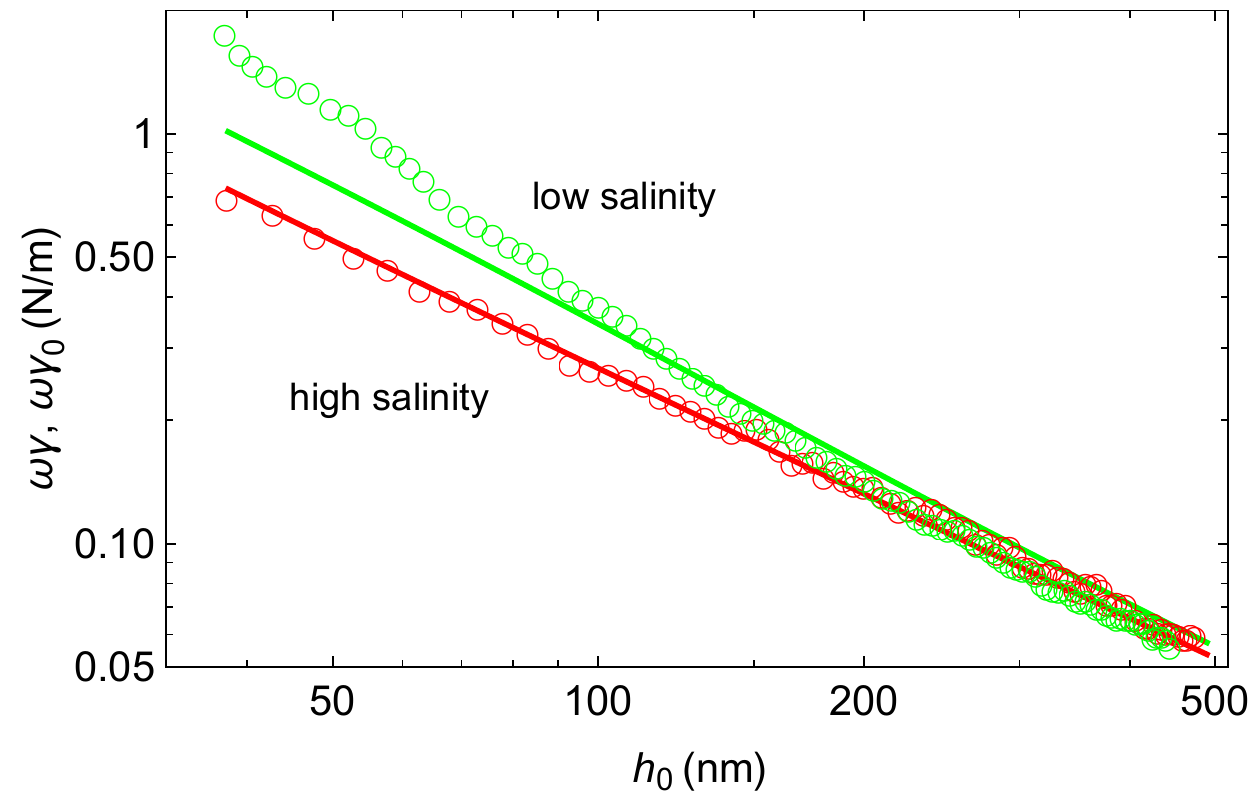}
	\caption{Comparison of the  drag coefficient measured at $\omega/2\pi=100$Hz (circles) with theory (solid curves). At high salinity ($\lambda<1$ nm, red), the data are well fitted by $\omega\gamma_0$  with the drag coefficient given by \eqref{eq:5}. At low salinity ($\lambda=47$ nm, green) we observe a significant  electroviscous enhancement, which is qualitatively accounted for by $\omega\gamma$ calculated from \eqref{eq:26}. For narrow gaps the measured data exceed the theory curve by up to 60$\%$.}
	\label{fig:F_exp} 
\end{figure}

If the double layers on either side of the water film don't overlap, their properties are given by the Poisson-Boltzmann potential \eqref{eq:13} calculated for an infinite half-space. As the surfaces get closer, the diffuse layers start to interact, resulting in electrostatic repulsion and electroviscous coupling. In the range where the distance $h_0$ is moderately larger than the Debye length $\lambda$, widely used approximations result in an exponentially screened electrostatic repulsion \cite{Israelachvili1986} and in a power-law dependence of the electroviscous drag \cite{Bike1990}.

Its range of validity is obviously related to the Debye length $\lambda$, yet our analysis shows that in reality it is limited by a significantly larger 
distance $\lambda_*$, defined in \eqref{eq:21}. With typical values of the $\zeta$-potential ranging from 25 to 100 mV, the parameter 
$\lambda_*$ may be up to 10 times larger than the actual screening length $\lambda$. This is clearly displayed by the electroviscous coupling parameter plotted in Fig. \ref{fig:xi}. 
The wide-channel approximation \label{eq3s2} converges 
only at $h_0\gg\lambda$. As a consequence, at distances of the order of or smaller than $\lambda_*$, the 
force can be calculated only numerically. 

\subsection{The effect of charge regulation}

\begin{figure}[tb]
	\includegraphics[width=\columnwidth]{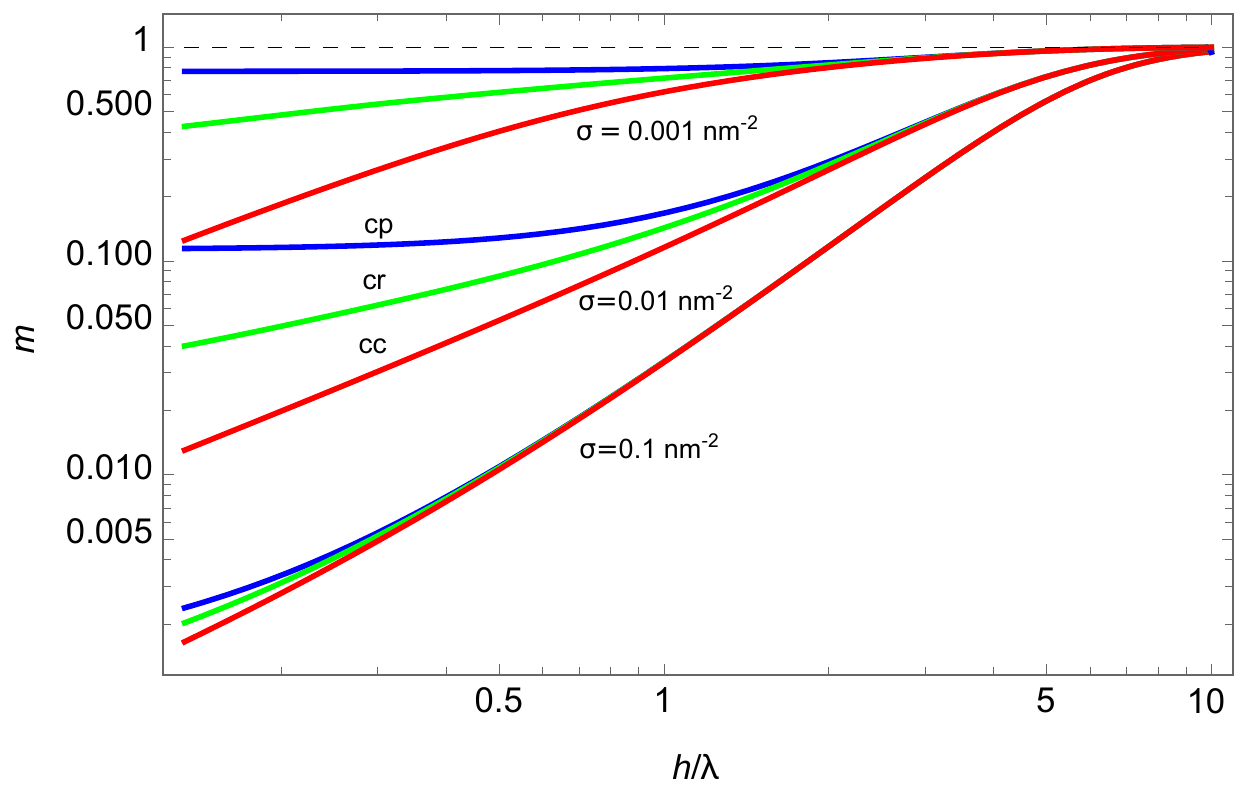}
	\caption{The parameter $m$ of the electrostatic potential \eqref{eq:48} as a function of reduced channel height $h/\lambda$, for three values of the surface charge density $\sigma$, and for constant charge (cc, red), constant potential (cp, blue), and charge regulation (cr, green). There are two different length scales: The onset of electroscatic coupling of the two diffuse layers, where $ m$ starts to decrease below 1, occurs at a distance $h_*=2\pi\sigma\ell_B\lambda^2$ which increases with $\sigma$. On the other hand, the electrostatic boundary condition and charge regulation (cc,cr,cp) are relevant at shorter distances, and there onset occurs at a distance which is inversely proportional to the surface charge density.} 
	\label{fig:parameter_m}
\end{figure}

There are two length scales indicating a qualitative change of the electrostatic properties, as illustrated by the parameter $m$ of the Jacobi elliptic function $\mathrm{cd}(u|m^2)$ in eq. \eqref{eq:48}, which is plotted in Fig. \ref{fig:parameter_m}. For very large channels one has $m=1$, which means that the double layers at opposite surfaces don't interact. The onset of the electrostatic coupling occurs at a film height $\lambda_*$ which increases with the surface charge density $\sigma$, as shown by the curves of Fig. \ref{fig:parameter_m}.

On the other hand, the electrostatic boundary conditions and charge regulation are relevant at smaller distances, and their onset shows the opposite behavior as a function of the surface charge density. Indeed, for $\sigma=0.001\mathrm{nm^{-2}}$ the three curves (cc, cp, cr) start diverging at $h=\lambda$, whereas for $\sigma=0.1\mathrm{nm^{-2}}$ this occurs at much smaller distances.

These features can be observed for both the electrostatic repulsion and electroviscous effects. Regarding the former, the two length scales for the onset of non-exponential behavior and charge regulation effects, are clearly visible in the inset of Fig. \ref{fig:staticforce_lin}. Similarly,   the  electroviscous  coupling  parameter $\xi$ in  Fig. \ref{fig:xi_CR} and the the enhancement of the drag coefficient 
in Fig.	\ref{fig:F_CR} show characteristic wide-channel power laws for $h\gg\lambda$, whereas charge regulation effects occur at distances shorter than the screening length.

{}{
\subsection{Electrokinetic lift force}
In this work we have considered the electroviscous force \eqref{eq:25} only. As pointed out by Bike and Prieve \cite{Bike1990}, there is an additional electrokinetic force, given by the diagonal part $\frac{1}{2}\varepsilon E^2$ of the Maxwell stress tensor, 
\be
F_\mathrm{el} = 2\pi R \int_{h_0}^\infty dh \frac{\varepsilon E^2}{2},
\ee
with the electric field \eqref{eq:17}. Because of $E\propto V$, this ``lift force'' is quadratic in the driving velocity $V\propto \cos\omega t$. As a consequence,  $F_\mathrm{el}\propto \cos^2\omega t$ is always repulsive and oscillates with the double frequency, contrary to the electroviscous force $F=-\gamma V$, which is opposite to the velocity and oscillates with $\omega$. }

{}{
The present experiments on squeezing motion do not show any indication of the lift force $F_\mathrm{el}$. This does not come as a surprise: inserting the wide-channel expressions of the transport coefficients $L_{ij}$ and a typical velocity $V=100$ nm/s, we find 
\be
F_\mathrm{el} \sim \varepsilon \zeta^2 
   \(\frac{ \lambda^2  }{ h_0^2 }\frac{ a\eta V R }{ k_BT }\)^2 \sim 10^{-17} \mathrm{N},
\ee
which is much smaller than the electroviscous force $F\sim 10^{-9} \mathrm{N}$. }

{}{
 For sliding motion along the surface, on the contrary, the lift force $F_\mathrm{el}$ turns out to be important. Due to the symmetry properties of the unperturbed pressure $P_0$, the corresponding  vertical force vanishes, $F=\int dS P_0=0$ \cite{Bike1990}. Moreover, the horizontal speed $\dot X$ of the sliding motion is typically of the order of 10 mm/s, much larger than the vertical velocity $V=\dot Z$ in the present experiment.}

\subsection{Comparison with previous work}
\label{nonlinear}

Electroviscous effects on squeezing motion have 
been studied in several previous papers \cite{Bike1990,Chun2004,Liu2018,Zhao2020}. All 
of these works start, more or less explicitly, 
from the volume and charge currents \eqref{eq:4} 
and \eqref{eq11s1}. Yet when calculating the charge current $J_C$, they use the unperturbed pressure gradient 
$\nabla P_0=-6\eta rV/h^3$ instead of 
$\nabla P$. This perturbative approach corresponds 
to a linearization of the pressure gradient in 
the coupling parameter $\xi$, 
\begin{align}
 \nabla P_1 = \nabla P_0 (1 + \xi)  ,
 \label{eq:41}
\end{align}
instead of the exact expression \eqref{eq:14}.

As a consequence, electroviscous effects appear 
as an additive correction to the unperturbed drag 
force $F_0$. Thus the wide-channel force of Bike 
and Prieve \cite{Bike1990} is identical to the first 
two terms of \eqref{eq:152}, whereas our expression \eqref{eq150sc} corresponds to the full series in  
$\lambda_*/h_0$.   Similarly, the numerical 
calculations of Chun and Ladd \cite{Chun2004} 
and Zhao \textit{et al.} \cite{Zhao2020}, are done with 
the linearized pressure gradient $P_1$. 
 
\begin{figure}[t]
	\includegraphics[width=\columnwidth]{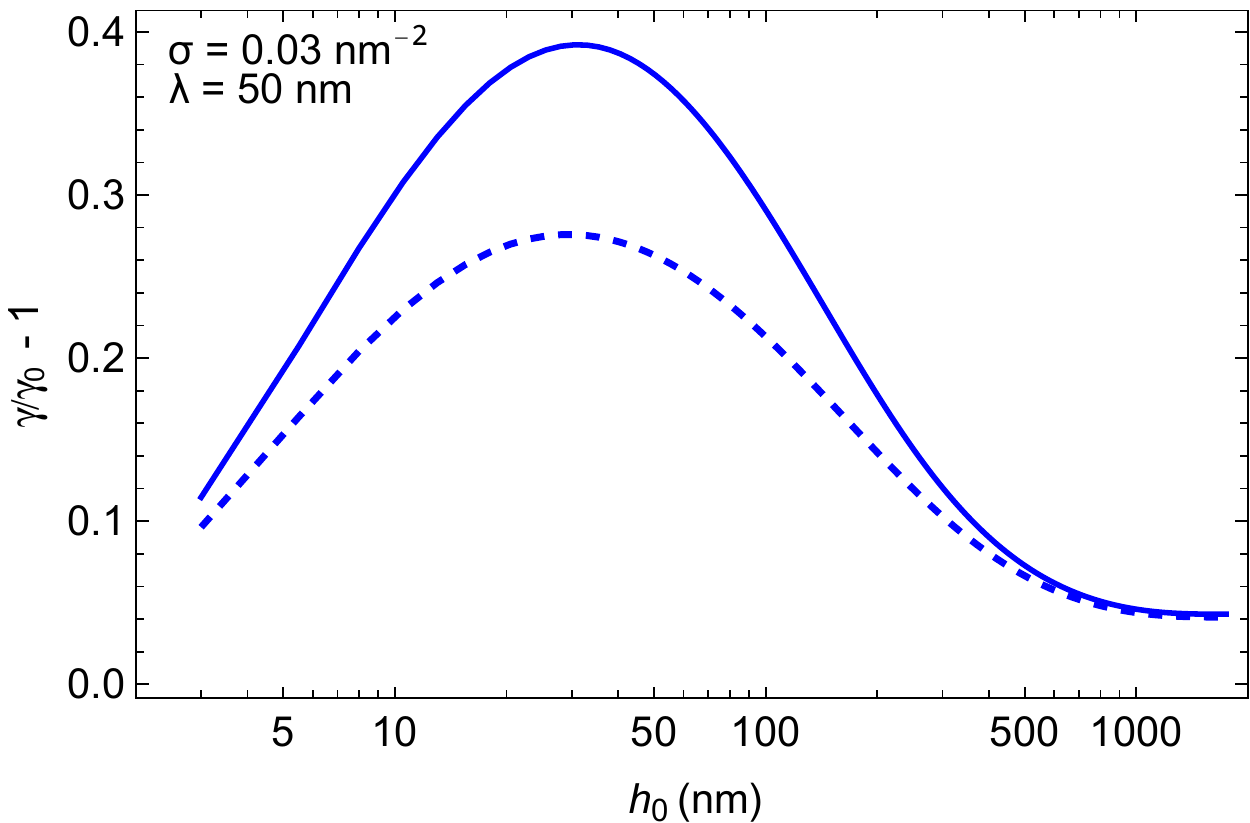}
	\caption{Numerical calculation of the electroviscous enhancement of the drag coefficient $\gamma/\gamma_0-1$ as a function of $h_0$, for $\sigma=0.03 \mathrm{nm}^{-2}$ and $\lambda=50$ nm.  The solid line is calculated with the full pressure \eqref{eq:14}, and the dashed line that with the linearized expression \eqref{eq:41}.}
	\label{fig:Flin}
\end{figure}

In Fig. \ref{fig:Flin} we compare the electroviscous 
enhancement of the drag force, calculated with the 
numerically exact pressure gradient \eqref{eq:14} 
and with the linearized form  $P_1$. For the parameters 
$\lambda=50\,\mathrm{nm}$ and $\sigma=0.03\,\mathrm{nm}^{-2}$, 
the linearized drag coefficient (dashed line) is by $28\%$ larger than $\gamma_0$, whereas the increase of the full expression (solid line) attains $40\%$. This difference is not surprising in view of the coupling parameters shown in Fig.~\ref{fig:xi}; in the intermediate range where $\xi$ reaches values of the order of unity, one expects a significant nonlinear behavior. 

\section{Summary}

We have studied the electroviscous and electrostatic forces exerted on a vibrating AFM tip across a nanoscale water film. We briefly summarize the main findings. 

{}{
(i) In the framework of Onsager relations for generalized fluxes and forces, we derive the drag coefficient \eqref{eq:26} in terms of the electroviscous coupling parameter $\xi$. With the surface charge $\sigma$ and the screening length $\lambda$ taken from the electrostatic repulsion (Fig. \ref{fig:static_force}), we find an almost quantitative agreement with experimental data (Fig. \ref{fig:F_exp}), with a discrepancy attaining 60$\%$ in the narrow-gap limit. 
}

(ii) This analysis relies on a quasistatic approximation \eqref{eq:15}, where the radial charge distribution in the water film is assumed to follow instantaneously the external driving. The fits of the viscous and elastic components of the response function \eqref{eq:70}, measured at $\omega/2\pi=100$ Hz and shown in Figs. \ref{fig:k} and \ref{fig:F_exp}, suggest that this approximation is justified at distances larger than the screening length, yet ceases to be valid for $h_0<\lambda$. Our experimental data strongly suggest that in this range both the spring constant $k$ and the drag coefficient $\gamma$ vary with frequency. The nature of the underlying relaxation process is not clear at present. 

(iii) Previous work relied on the linearization approximation \eqref{eq:41} for the hydrodynamic pressure gradient. This linearization significantly underestimates the enhancement of the drag coefficients, for the parameters of Fig. \ref{fig:Flin} by about 40 $\%$. 

(iv) Charge regulation turns out to be of minor importance in the experimentally most relevant range. Indeed, the electroviscous coupliing sets in at large distances and is maximum at $h_0\sim3\lambda$ (Fig. \ref{fig:xi}), whereas the electrostatic boundary conditions and charge regulation effects are significant in narrow channels only, as shown in Figs. \ref{fig:xi_CR}--\ref{fig:static_force}.

\section{Acknowledgements}

This project was supported by the French National 
Research Agency through Grant No. ANR-19-CE30-0012. 
MRM  acknowledges funding support from the Mexican 
Council for Science and Technology (CONACYT) Grant No. 
CVU 625862. ZZ acknowledges funding support from the 
China Scholarship Council.

\section*{}
$^*$MRM, ZZ and ZB contributed equally to this work. 

\vspace{10pt}

\bibliography{biblio}

\begin{thebibliography}{37}%
\makeatletter
\providecommand \@ifxundefined [1]{%
 \@ifx{#1\undefined}
}%
\providecommand \@ifnum [1]{%
 \ifnum #1\expandafter \@firstoftwo
 \else \expandafter \@secondoftwo
 \fi
}%
\providecommand \@ifx [1]{%
 \ifx #1\expandafter \@firstoftwo
 \else \expandafter \@secondoftwo
 \fi
}%
\providecommand \natexlab [1]{#1}%
\providecommand \enquote  [1]{``#1''}%
\providecommand \bibnamefont  [1]{#1}%
\providecommand \bibfnamefont [1]{#1}%
\providecommand \citenamefont [1]{#1}%
\providecommand \href@noop [0]{\@secondoftwo}%
\providecommand \href [0]{\begingroup \@sanitize@url \@href}%
\providecommand \@href[1]{\@@startlink{#1}\@@href}%
\providecommand \@@href[1]{\endgroup#1\@@endlink}%
\providecommand \@sanitize@url [0]{\catcode `\\12\catcode `\$12\catcode
  `\&12\catcode `\#12\catcode `\^12\catcode `\_12\catcode `\%12\relax}%
\providecommand \@@startlink[1]{}%
\providecommand \@@endlink[0]{}%
\providecommand \url  [0]{\begingroup\@sanitize@url \@url }%
\providecommand \@url [1]{\endgroup\@href {#1}{\urlprefix }}%
\providecommand \urlprefix  [0]{URL }%
\providecommand \Eprint [0]{\href }%
\providecommand \doibase [0]{http://dx.doi.org/}%
\providecommand \selectlanguage [0]{\@gobble}%
\providecommand \bibinfo  [0]{\@secondoftwo}%
\providecommand \bibfield  [0]{\@secondoftwo}%
\providecommand \translation [1]{[#1]}%
\providecommand \BibitemOpen [0]{}%
\providecommand \bibitemStop [0]{}%
\providecommand \bibitemNoStop [0]{.\EOS\space}%
\providecommand \EOS [0]{\spacefactor3000\relax}%
\providecommand \BibitemShut  [1]{\csname bibitem#1\endcsname}%
\let\auto@bib@innerbib\@empty
%</preamble>
\bibitem [{\citenamefont {Israelachvili}()}]{Israelachvili1991}%
  \BibitemOpen
  \bibfield  {author} {\bibinfo {author} {\bibfnamefont {J.~N.}\ \bibnamefont
  {Israelachvili}},\ }\href@noop {} {\emph {\bibinfo {title} {Intermolecular
  and surface forces}}},\ \bibinfo {edition} {2nd}\ ed.\ (\bibinfo  {publisher}
  {Academic Press London, San Diego, 1991})\BibitemShut {NoStop}%
\bibitem [{\citenamefont {Israelachvili}\ and\ \citenamefont
  {Wennerstr{\"o}m}(1996)}]{israelachvili1996role}%
  \BibitemOpen
  \bibfield  {author} {\bibinfo {author} {\bibfnamefont {J.}~\bibnamefont
  {Israelachvili}}\ and\ \bibinfo {author} {\bibfnamefont {H.}~\bibnamefont
  {Wennerstr{\"o}m}},\ }\href@noop {} {\bibfield  {journal} {\bibinfo
  {journal} {Nature}\ }\textbf {\bibinfo {volume} {379}},\ \bibinfo {pages}
  {219} (\bibinfo {year} {1996})}\BibitemShut {NoStop}%
\bibitem [{\citenamefont {Lyklema}(1995)}]{Lyklema1995}%
  \BibitemOpen
  \bibfield  {author} {\bibinfo {author} {\bibfnamefont {J.}~\bibnamefont
  {Lyklema}},\ }\href@noop {} {\emph {\bibinfo {title} {Fundamentals of
  Interface and Colloid Science}}}\ (\bibinfo  {publisher} {Academic Press},\
  \bibinfo {address} {New York},\ \bibinfo {year} {1995})\BibitemShut {NoStop}%
\bibitem [{\citenamefont {Stone}\ \emph {et~al.}(2004)\citenamefont {Stone},
  \citenamefont {Stroock},\ and\ \citenamefont
  {Ajdari}}]{stone2004engineering}%
  \BibitemOpen
  \bibfield  {author} {\bibinfo {author} {\bibfnamefont {H.~A.}\ \bibnamefont
  {Stone}}, \bibinfo {author} {\bibfnamefont {A.~D.}\ \bibnamefont {Stroock}},
  \ and\ \bibinfo {author} {\bibfnamefont {A.}~\bibnamefont {Ajdari}},\
  }\href@noop {} {\bibfield  {journal} {\bibinfo  {journal} {Annu. Rev. Fluid
  Mech.}\ }\textbf {\bibinfo {volume} {36}},\ \bibinfo {pages} {381} (\bibinfo
  {year} {2004})}\BibitemShut {NoStop}%
\bibitem [{\citenamefont {Yan}\ \emph {et~al.}(2016)\citenamefont {Yan},
  \citenamefont {Han}, \citenamefont {Zhang}, \citenamefont {Xu}, \citenamefont
  {Luijten},\ and\ \citenamefont {Granick}}]{yan2016reconfiguring}%
  \BibitemOpen
  \bibfield  {author} {\bibinfo {author} {\bibfnamefont {J.}~\bibnamefont
  {Yan}}, \bibinfo {author} {\bibfnamefont {M.}~\bibnamefont {Han}}, \bibinfo
  {author} {\bibfnamefont {J.}~\bibnamefont {Zhang}}, \bibinfo {author}
  {\bibfnamefont {C.}~\bibnamefont {Xu}}, \bibinfo {author} {\bibfnamefont
  {E.}~\bibnamefont {Luijten}}, \ and\ \bibinfo {author} {\bibfnamefont
  {S.}~\bibnamefont {Granick}},\ }\href@noop {} {\bibfield  {journal} {\bibinfo
   {journal} {Nature materials}\ }\textbf {\bibinfo {volume} {15}},\ \bibinfo
  {pages} {1095} (\bibinfo {year} {2016})}\BibitemShut {NoStop}%
\bibitem [{\citenamefont {Marbach}\ and\ \citenamefont
  {Bocquet}(2019)}]{Marbach2019}%
  \BibitemOpen
  \bibfield  {author} {\bibinfo {author} {\bibfnamefont {S.}~\bibnamefont
  {Marbach}}\ and\ \bibinfo {author} {\bibfnamefont {L.}~\bibnamefont
  {Bocquet}},\ }\href@noop {} {\bibfield  {journal} {\bibinfo  {journal} {Chem.
  Soc. Rev.}\ }\textbf {\bibinfo {volume} {48}},\ \bibinfo {pages} {3102}
  (\bibinfo {year} {2019})}\BibitemShut {NoStop}%
\bibitem [{\citenamefont {Bocquet}\ and\ \citenamefont
  {Charlaix}(2010)}]{bocquet2010nanofluidics}%
  \BibitemOpen
  \bibfield  {author} {\bibinfo {author} {\bibfnamefont {L.}~\bibnamefont
  {Bocquet}}\ and\ \bibinfo {author} {\bibfnamefont {E.}~\bibnamefont
  {Charlaix}},\ }\href@noop {} {\bibfield  {journal} {\bibinfo  {journal}
  {Chemical Reviews}\ }\textbf {\bibinfo {volume} {39}},\ \bibinfo {pages}
  {1073} (\bibinfo {year} {2010})}\BibitemShut {NoStop}%
\bibitem [{\citenamefont {Helmholtz}(1879)}]{Helmholtz1879}%
  \BibitemOpen
  \bibfield  {author} {\bibinfo {author} {\bibfnamefont {H.}~\bibnamefont
  {Helmholtz}},\ }\href {\doibase https://doi.org/10.1002/andp.18792430702}
  {\bibfield  {journal} {\bibinfo  {journal} {Annalen der Physik}\ }\textbf
  {\bibinfo {volume} {243}},\ \bibinfo {pages} {337} (\bibinfo {year}
  {1879})}\BibitemShut {NoStop}%
\bibitem [{\citenamefont {von Smoluchowski}(1903)}]{Smoluchowski1903}%
  \BibitemOpen
  \bibfield  {author} {\bibinfo {author} {\bibfnamefont {M.}~\bibnamefont {von
  Smoluchowski}},\ }\href@noop {} {\bibfield  {journal} {\bibinfo  {journal}
  {Bull. Akad. Sci. Cracovie.}\ }\textbf {\bibinfo {volume} {8}},\ \bibinfo
  {pages} {182} (\bibinfo {year} {1903})}\BibitemShut {NoStop}%
\bibitem [{\citenamefont {Bikerman}(1933)}]{Bikerman1933}%
  \BibitemOpen
  \bibfield  {author} {\bibinfo {author} {\bibfnamefont {J.~J.}\ \bibnamefont
  {Bikerman}},\ }\href@noop {} {\bibfield  {journal} {\bibinfo  {journal}
  {Zeitschrift f{\"u}r Elektrochemie und angewandte physikalische Chemie}\
  }\textbf {\bibinfo {volume} {39}},\ \bibinfo {pages} {526} (\bibinfo {year}
  {1933})}\BibitemShut {NoStop}%
\bibitem [{\citenamefont {H\"uckel}(1924)}]{Hueckel1924}%
  \BibitemOpen
  \bibfield  {author} {\bibinfo {author} {\bibfnamefont {E.}~\bibnamefont
  {H\"uckel}},\ }\href@noop {} {\bibfield  {journal} {\bibinfo  {journal}
  {Physik. Z.}\ }\textbf {\bibinfo {volume} {25}},\ \bibinfo {pages} {97}
  (\bibinfo {year} {1924})}\BibitemShut {NoStop}%
\bibitem [{\citenamefont {Henry}\ and\ \citenamefont
  {Lapworth}(1931)}]{Henry1931}%
  \BibitemOpen
  \bibfield  {author} {\bibinfo {author} {\bibfnamefont {D.~C.}\ \bibnamefont
  {Henry}}\ and\ \bibinfo {author} {\bibfnamefont {A.}~\bibnamefont
  {Lapworth}},\ }\href@noop {} {\bibfield  {journal} {\bibinfo  {journal}
  {Proceedings of the Royal Society of London. Series A}\ }\textbf {\bibinfo
  {volume} {133}},\ \bibinfo {pages} {106} (\bibinfo {year}
  {1931})}\BibitemShut {NoStop}%
\bibitem [{\citenamefont {Gross}\ and\ \citenamefont
  {Osterle}(1968)}]{Gross1968}%
  \BibitemOpen
  \bibfield  {author} {\bibinfo {author} {\bibfnamefont {R.~J.}\ \bibnamefont
  {Gross}}\ and\ \bibinfo {author} {\bibfnamefont {J.~F.}\ \bibnamefont
  {Osterle}},\ }\href {\doibase 10.1063/1.1669814} {\bibfield  {journal}
  {\bibinfo  {journal} {The Journal of Chemical Physics}\ }\textbf {\bibinfo
  {volume} {49}},\ \bibinfo {pages} {228} (\bibinfo {year} {1968})}\BibitemShut
  {NoStop}%
\bibitem [{\citenamefont {Alexander}\ and\ \citenamefont
  {Prieve}(1987)}]{Alexander1987}%
  \BibitemOpen
  \bibfield  {author} {\bibinfo {author} {\bibfnamefont {B.~M.}\ \bibnamefont
  {Alexander}}\ and\ \bibinfo {author} {\bibfnamefont {D.~C.}\ \bibnamefont
  {Prieve}},\ }\href {\doibase 10.1021/la00077a038} {\bibfield  {journal}
  {\bibinfo  {journal} {Langmuir}\ }\textbf {\bibinfo {volume} {3}},\ \bibinfo
  {pages} {788} (\bibinfo {year} {1987})}\BibitemShut {NoStop}%
\bibitem [{\citenamefont {Bike}\ and\ \citenamefont {Prieve}(1990)}]{Bike1990}%
  \BibitemOpen
  \bibfield  {author} {\bibinfo {author} {\bibfnamefont {S.~G.}\ \bibnamefont
  {Bike}}\ and\ \bibinfo {author} {\bibfnamefont {D.~C.}\ \bibnamefont
  {Prieve}},\ }\href@noop {} {\bibfield  {journal} {\bibinfo  {journal}
  {Journal of Colloid and Interface Science}\ }\textbf {\bibinfo {volume}
  {136}},\ \bibinfo {pages} {95} (\bibinfo {year} {1990})}\BibitemShut
  {NoStop}%
\bibitem [{\citenamefont {Bike}\ and\ \citenamefont {Prieve}(1992)}]{Bike1992}%
  \BibitemOpen
  \bibfield  {author} {\bibinfo {author} {\bibfnamefont {S.~G.}\ \bibnamefont
  {Bike}}\ and\ \bibinfo {author} {\bibfnamefont {D.~C.}\ \bibnamefont
  {Prieve}},\ }\href {\doibase https://doi.org/10.1016/0021-9797(92)90080-6}
  {\bibfield  {journal} {\bibinfo  {journal} {Journal of Colloid and Interface
  Science}\ }\textbf {\bibinfo {volume} {154}},\ \bibinfo {pages} {87}
  (\bibinfo {year} {1992})}\BibitemShut {NoStop}%
\bibitem [{\citenamefont {Bike}\ \emph {et~al.}(1995)\citenamefont {Bike},
  \citenamefont {Lazarro},\ and\ \citenamefont {Prieve}}]{Bike1995a}%
  \BibitemOpen
  \bibfield  {author} {\bibinfo {author} {\bibfnamefont {S.~G.}\ \bibnamefont
  {Bike}}, \bibinfo {author} {\bibfnamefont {L.}~\bibnamefont {Lazarro}}, \
  and\ \bibinfo {author} {\bibfnamefont {D.~C.}\ \bibnamefont {Prieve}},\
  }\href {\doibase https://doi.org/10.1006/jcis.1995.1471} {\bibfield
  {journal} {\bibinfo  {journal} {Journal of Colloid and Interface Science}\
  }\textbf {\bibinfo {volume} {175}},\ \bibinfo {pages} {411} (\bibinfo {year}
  {1995})}\BibitemShut {NoStop}%
\bibitem [{\citenamefont {Bike}\ and\ \citenamefont
  {Prieve}(1995)}]{Bike1995b}%
  \BibitemOpen
  \bibfield  {author} {\bibinfo {author} {\bibfnamefont {S.~G.}\ \bibnamefont
  {Bike}}\ and\ \bibinfo {author} {\bibfnamefont {D.~C.}\ \bibnamefont
  {Prieve}},\ }\href {\doibase https://doi.org/10.1006/jcis.1995.1472}
  {\bibfield  {journal} {\bibinfo  {journal} {Journal of Colloid and Interface
  Science}\ }\textbf {\bibinfo {volume} {175}},\ \bibinfo {pages} {422}
  (\bibinfo {year} {1995})}\BibitemShut {NoStop}%
\bibitem [{\citenamefont {Raiteri}\ \emph {et~al.}(1996)\citenamefont
  {Raiteri}, \citenamefont {Grattarola},\ and\ \citenamefont
  {Butt}}]{raiteri1996measuring}%
  \BibitemOpen
  \bibfield  {author} {\bibinfo {author} {\bibfnamefont {R.}~\bibnamefont
  {Raiteri}}, \bibinfo {author} {\bibfnamefont {M.}~\bibnamefont {Grattarola}},
  \ and\ \bibinfo {author} {\bibfnamefont {H.-J.}\ \bibnamefont {Butt}},\
  }\href@noop {} {\bibfield  {journal} {\bibinfo  {journal} {The Journal of
  Physical Chemistry}\ }\textbf {\bibinfo {volume} {100}},\ \bibinfo {pages}
  {16700} (\bibinfo {year} {1996})}\BibitemShut {NoStop}%
\bibitem [{\citenamefont {Liu}\ \emph {et~al.}(2015)\citenamefont {Liu},
  \citenamefont {Zhao}, \citenamefont {Mugele},\ and\ \citenamefont {van~den
  Ende}}]{Liu2015}%
  \BibitemOpen
  \bibfield  {author} {\bibinfo {author} {\bibfnamefont {F.}~\bibnamefont
  {Liu}}, \bibinfo {author} {\bibfnamefont {C.}~\bibnamefont {Zhao}}, \bibinfo
  {author} {\bibfnamefont {F.}~\bibnamefont {Mugele}}, \ and\ \bibinfo {author}
  {\bibfnamefont {D.}~\bibnamefont {van~den Ende}},\ }\href {\doibase
  10.1088/0957-4484/26/38/385703} {\bibfield  {journal} {\bibinfo  {journal}
  {Nanotechnology}\ }\textbf {\bibinfo {volume} {26}},\ \bibinfo {pages}
  {385703} (\bibinfo {year} {2015})}\BibitemShut {NoStop}%
\bibitem [{\citenamefont {Liu}\ \emph {et~al.}(2018)\citenamefont {Liu},
  \citenamefont {Klaassen}, \citenamefont {Zhao}, \citenamefont {Mugele},\ and\
  \citenamefont {van~den Ende}}]{Liu2018}%
  \BibitemOpen
  \bibfield  {author} {\bibinfo {author} {\bibfnamefont {F.}~\bibnamefont
  {Liu}}, \bibinfo {author} {\bibfnamefont {A.}~\bibnamefont {Klaassen}},
  \bibinfo {author} {\bibfnamefont {C.}~\bibnamefont {Zhao}}, \bibinfo {author}
  {\bibfnamefont {F.}~\bibnamefont {Mugele}}, \ and\ \bibinfo {author}
  {\bibfnamefont {D.}~\bibnamefont {van~den Ende}},\ }\href {\doibase
  10.1021/acs.jpcb.7b07019} {\bibfield  {journal} {\bibinfo  {journal} {The
  Journal of Physical Chemistry B}\ }\textbf {\bibinfo {volume} {122}},\
  \bibinfo {pages} {933} (\bibinfo {year} {2018})},\ \bibinfo {note} {pMID:
  28976197}\BibitemShut {NoStop}%
\bibitem [{\citenamefont {Chun}\ and\ \citenamefont {Ladd}(2004)}]{Chun2004}%
  \BibitemOpen
  \bibfield  {author} {\bibinfo {author} {\bibfnamefont {B.}~\bibnamefont
  {Chun}}\ and\ \bibinfo {author} {\bibfnamefont {A.}~\bibnamefont {Ladd}},\
  }\href {\doibase https://doi.org/10.1016/j.jcis.2004.03.066} {\bibfield
  {journal} {\bibinfo  {journal} {Journal of Colloid and Interface Science}\
  }\textbf {\bibinfo {volume} {274}},\ \bibinfo {pages} {687} (\bibinfo {year}
  {2004})}\BibitemShut {NoStop}%
\bibitem [{\citenamefont {Zhao}\ \emph {et~al.}(2020)\citenamefont {Zhao},
  \citenamefont {Zhang}, \citenamefont {van~den Ende},\ and\ \citenamefont
  {Mugele}}]{Zhao2020}%
  \BibitemOpen
  \bibfield  {author} {\bibinfo {author} {\bibfnamefont {C.}~\bibnamefont
  {Zhao}}, \bibinfo {author} {\bibfnamefont {W.}~\bibnamefont {Zhang}},
  \bibinfo {author} {\bibfnamefont {D.}~\bibnamefont {van~den Ende}}, \ and\
  \bibinfo {author} {\bibfnamefont {F.}~\bibnamefont {Mugele}},\ }\href
  {\doibase 10.1017/jfm.2020.68} {\bibfield  {journal} {\bibinfo  {journal}
  {Journal of Fluid Mechanics}\ }\textbf {\bibinfo {volume} {888}},\ \bibinfo
  {pages} {A29} (\bibinfo {year} {2020})}\BibitemShut {NoStop}%
\bibitem [{\citenamefont {Andelman}(2006)}]{Andelman2006}%
  \BibitemOpen
  \bibfield  {author} {\bibinfo {author} {\bibfnamefont {D.}~\bibnamefont
  {Andelman}},\ }in\ \href@noop {} {\emph {\bibinfo {booktitle} {Soft Condensed
  Matter Physics in Molecular and Cell Biology}}},\ \bibinfo {editor} {edited
  by\ \bibinfo {editor} {\bibfnamefont {W.}~\bibnamefont {Poon}}\ and\ \bibinfo
  {editor} {\bibfnamefont {D.}~\bibnamefont {Andelman}}}\ (\bibinfo
  {publisher} {Taylor $\&$ Francis},\ \bibinfo {address} {New York},\ \bibinfo
  {year} {2006})\ pp.\ \bibinfo {pages} {97--122}\BibitemShut {NoStop}%
\bibitem [{\citenamefont {Derjaguin}(1934)}]{Derjaguin1934}%
  \BibitemOpen
  \bibfield  {author} {\bibinfo {author} {\bibfnamefont {B.}~\bibnamefont
  {Derjaguin}},\ }\href {\doibase 10.1007/BF01433225} {\bibfield  {journal}
  {\bibinfo  {journal} {Kolloid-Zeitschrift}\ }\textbf {\bibinfo {volume}
  {69}},\ \bibinfo {pages} {155} (\bibinfo {year} {1934})}\BibitemShut
  {NoStop}%
\bibitem [{\citenamefont {Happel}\ and\ \citenamefont
  {Brenner}(1963)}]{Happel1963}%
  \BibitemOpen
  \bibfield  {author} {\bibinfo {author} {\bibfnamefont {J.}~\bibnamefont
  {Happel}}\ and\ \bibinfo {author} {\bibfnamefont {H.}~\bibnamefont
  {Brenner}},\ }\href@noop {} {\emph {\bibinfo {title} {Low Reynolds Number
  Hydrodynamics}}}\ (\bibinfo  {publisher} {Martinus Nijhoff},\ \bibinfo {year}
  {1963})\BibitemShut {NoStop}%
\bibitem [{\citenamefont {Anderson}(1989)}]{Anderson1989}%
  \BibitemOpen
  \bibfield  {author} {\bibinfo {author} {\bibfnamefont {J.~L.}\ \bibnamefont
  {Anderson}},\ }\href@noop {} {\bibfield  {journal} {\bibinfo  {journal}
  {Annu. Rev. Fluid Mech.}\ }\textbf {\bibinfo {volume} {21}},\ \bibinfo
  {pages} {61} (\bibinfo {year} {1989})}\BibitemShut {NoStop}%
\bibitem [{\citenamefont {Brenner}(1961)}]{Brenner1961}%
  \BibitemOpen
  \bibfield  {author} {\bibinfo {author} {\bibfnamefont {H.}~\bibnamefont
  {Brenner}},\ }\href@noop {} {\bibfield  {journal} {\bibinfo  {journal} {Chem.
  Eng. Sci.}\ }\textbf {\bibinfo {volume} {16}},\ \bibinfo {pages} {242}
  (\bibinfo {year} {1961})}\BibitemShut {NoStop}%
\bibitem [{\citenamefont {Stein}\ \emph {et~al.}(2004)\citenamefont {Stein},
  \citenamefont {Kruithof},\ and\ \citenamefont {Dekker}}]{stein2004surface}%
  \BibitemOpen
  \bibfield  {author} {\bibinfo {author} {\bibfnamefont {D.}~\bibnamefont
  {Stein}}, \bibinfo {author} {\bibfnamefont {M.}~\bibnamefont {Kruithof}}, \
  and\ \bibinfo {author} {\bibfnamefont {C.}~\bibnamefont {Dekker}},\
  }\href@noop {} {\bibfield  {journal} {\bibinfo  {journal} {Physical Review
  Letters}\ }\textbf {\bibinfo {volume} {93}},\ \bibinfo {pages} {035901}
  (\bibinfo {year} {2004})}\BibitemShut {NoStop}%
\bibitem [{\citenamefont {{Abramowitz}}\ and\ \citenamefont
  {{Stegun}}(1964)}]{Abramowitz1964}%
  \BibitemOpen
  \bibfield  {author} {\bibinfo {author} {\bibfnamefont {M.}~\bibnamefont
  {{Abramowitz}}}\ and\ \bibinfo {author} {\bibfnamefont {I.~A.}\ \bibnamefont
  {{Stegun}}},\ }\href@noop {} {\emph {\bibinfo {title} {Handbook of
  Mathematical Functions}}}\ (\bibinfo  {publisher} {Dover},\ \bibinfo
  {address} {New York},\ \bibinfo {year} {1964})\BibitemShut {NoStop}%
\bibitem [{\citenamefont {Bulavin}\ \emph {et~al.}(2011)\citenamefont
  {Bulavin}, \citenamefont {Zhyganiuk}, \citenamefont {Malomuzh},\ and\
  \citenamefont {Pankratov}}]{Bulavin2011}%
  \BibitemOpen
  \bibfield  {author} {\bibinfo {author} {\bibfnamefont {L.}~\bibnamefont
  {Bulavin}}, \bibinfo {author} {\bibfnamefont {I.}~\bibnamefont {Zhyganiuk}},
  \bibinfo {author} {\bibfnamefont {M.}~\bibnamefont {Malomuzh}}, \ and\
  \bibinfo {author} {\bibfnamefont {K.}~\bibnamefont {Pankratov}},\ }\href@noop
  {} {\bibfield  {journal} {\bibinfo  {journal} {Ukr. J. Phys.}\ }\textbf
  {\bibinfo {volume} {56}},\ \bibinfo {pages} {893} (\bibinfo {year}
  {2011})}\BibitemShut {NoStop}%
\bibitem [{\citenamefont {Ninham}\ and\ \citenamefont
  {Parsegian}(1971)}]{Nin71}%
  \BibitemOpen
  \bibfield  {author} {\bibinfo {author} {\bibfnamefont {B.~W.}\ \bibnamefont
  {Ninham}}\ and\ \bibinfo {author} {\bibfnamefont {V.~A.}\ \bibnamefont
  {Parsegian}},\ }\href {\doibase https://doi.org/10.1016/0022-5193(71)90019-1}
  {\bibfield  {journal} {\bibinfo  {journal} {J. Theor. Biol.}\ }\textbf
  {\bibinfo {volume} {31}},\ \bibinfo {pages} {405} (\bibinfo {year}
  {1971})}\BibitemShut {NoStop}%
\bibitem [{\citenamefont {Majee}\ \emph {et~al.}(2018)\citenamefont {Majee},
  \citenamefont {Bier},\ and\ \citenamefont {Podgornik}}]{Maj18}%
  \BibitemOpen
  \bibfield  {author} {\bibinfo {author} {\bibfnamefont {A.}~\bibnamefont
  {Majee}}, \bibinfo {author} {\bibfnamefont {M.}~\bibnamefont {Bier}}, \ and\
  \bibinfo {author} {\bibfnamefont {R.}~\bibnamefont {Podgornik}},\ }\href
  {\doibase 10.1039/c7sm02270k} {\bibfield  {journal} {\bibinfo  {journal}
  {Soft Matter}\ }\textbf {\bibinfo {volume} {14}},\ \bibinfo {pages} {985}
  (\bibinfo {year} {2018})}\BibitemShut {NoStop}%
\bibitem [{\citenamefont {Markovich}\ \emph {et~al.}(2016)\citenamefont
  {Markovich}, \citenamefont {Andelman},\ and\ \citenamefont
  {Podgornik}}]{Markovich_2016}%
  \BibitemOpen
  \bibfield  {author} {\bibinfo {author} {\bibfnamefont {T.}~\bibnamefont
  {Markovich}}, \bibinfo {author} {\bibfnamefont {D.}~\bibnamefont {Andelman}},
  \ and\ \bibinfo {author} {\bibfnamefont {R.}~\bibnamefont {Podgornik}},\
  }\href@noop {} {\bibfield  {journal} {\bibinfo  {journal} {Europhys. Lett.}\
  }\textbf {\bibinfo {volume} {113}},\ \bibinfo {pages} {26004} (\bibinfo
  {year} {2016})}\BibitemShut {NoStop}%
\bibitem [{\citenamefont {Maali}\ and\ \citenamefont
  {Boisgard}(2013)}]{maali2013precise}%
  \BibitemOpen
  \bibfield  {author} {\bibinfo {author} {\bibfnamefont {A.}~\bibnamefont
  {Maali}}\ and\ \bibinfo {author} {\bibfnamefont {R.}~\bibnamefont
  {Boisgard}},\ }\href@noop {} {\bibfield  {journal} {\bibinfo  {journal}
  {Journal of Applied Physics}\ }\textbf {\bibinfo {volume} {114}},\ \bibinfo
  {pages} {144302} (\bibinfo {year} {2013})}\BibitemShut {NoStop}%
\bibitem [{\citenamefont {Craig}\ and\ \citenamefont {Neto}(2001)}]{Craig2001}%
  \BibitemOpen
  \bibfield  {author} {\bibinfo {author} {\bibfnamefont {V.~S.~J.}\
  \bibnamefont {Craig}}\ and\ \bibinfo {author} {\bibfnamefont
  {C.}~\bibnamefont {Neto}},\ }\href@noop {} {\bibfield  {journal} {\bibinfo
  {journal} {Langmuir}\ }\textbf {\bibinfo {volume} {17}},\ \bibinfo {pages}
  {6018} (\bibinfo {year} {2001})}\BibitemShut {NoStop}%
\bibitem [{\citenamefont {Israelachvili}(1986)}]{Israelachvili1986}%
  \BibitemOpen
  \bibfield  {author} {\bibinfo {author} {\bibfnamefont {J.}~\bibnamefont
  {Israelachvili}},\ }\href@noop {} {\bibfield  {journal} {\bibinfo  {journal}
  {Journal of Colloid and Interface Science}\ }\textbf {\bibinfo {volume}
  {110}},\ \bibinfo {pages} {263} (\bibinfo {year} {1986})}\BibitemShut
  {NoStop}%
\end{thebibliography}%

\end{document}